\documentclass[journal=jpcafh,manuscript=article]{achemso}
\setkeys{acs}{usetitle=true}
\usepackage{graphicx}
\usepackage{xcolor}
\usepackage{longtable}

\title{
  Entanglement Effect and Angular Momentum Conservation in a
  Non-separable Tunneling Treatment
}
\SectionNumbersOn

\author{Yuri Georgievskii}
\email{ygeorgi@anl.gov}
\author{Stephen J. Klippenstein}
\email{sjk@anl.gov}

\affiliation{ Chemical Sciences and Engineering Division, Argonne
National Laboratory, Lemont, IL 60439}

\begin{document}
\begin{abstract}
  The important, and often dominant, role of tunneling in low
  temperature kinetics has resulted in numerous theoretical
  explorations into the methodology for predicting it. Nevertheless,
  there are still key aspects of the derivations that are lacking,
  particularly for non-separable systems in the low temperature
  regime, and further explorations of the physical factors affecting
  the tunneling rate are warranted.  In this work we obtain a
  closed-form rate expression for the tunneling rate constant that is
  a direct analog of the rigid-rotor-harmonic-oscillator expression.
  This expression introduces a novel "entanglement factor" that
  modulates the reaction rate.  Furthermore, we are able to extend
  this expression, which is valid for non-separable systems at low
  temperatures, to properly account for the conservation of angular
  momentum.  In contrast, previous calculations have considered only
  vibrational transverse modes and so effectively employ a decoupled
  rotational partition function for the orientational modes.  We also
  suggest a simple theoretical model to describe the tunneling effects
  in the vicinity of the crossover temperature (the temperature where
  tunneling becomes the dominating mechanism). This model allows one
  to naturally classify, interpret, and predict experimental
  data. Among other things, it quantitatively explains in simple terms
  the so-called ``quantum bobsled'' effect, also known as the negative
  centrifugal effect, which is related to curvature of the reaction
  path. Taken together, the expressions obtained here allow one to
  predict the thermal and $E$-resolved rate constants over broad
  ranges of temperatures and energies.
\end{abstract}

\newcommand{\pint}{\ensuremath{\overbrace{\underbrace{\int\cdots\int}}\mathcal{D}}}

\renewcommand{\thesection}{\arabic{section}.}
\renewcommand{\thesubsection}{\arabic{section}.\arabic{subsection}}
\renewcommand{\theequation}{\arabic{section}.\arabic{equation}}

\setcounter{section}{0}
\setcounter{equation}{0}

\newpage
\section{Introduction}
 The famous transition state (TS) theory high pressure rate
constant expression reads\cite{glasstone41} (we use atomic units, $\hbar = 1$)
\begin{equation}
  \label{tst}
  k = \frac{k_BT}{2\pi}\frac{Z^\#}{Z_R} e^{-V^\#/ k_BT},
\end{equation}
where $T$ and $k_B$ are temperature and the Boltzmann constant, $Z_R$
is the partition function of the reactant(s), and $V^\#$ is the
barrier height. The partition function for the TS dividing surface
$Z^\#$ can be factorized into rotational and vibrational contributions
in the framework of the rigid-rotor harmonic-oscillator (RRHO)
approximation at the saddle point,
\begin{eqnarray}
  Z^\#       &=& Z^\#_{rot}Z^\#_{vib},
                 \label{rrho_notun}\\
  Z^\#_{rot }&=& 2\sqrt{2\pi}\sqrt{I_xI_yI_z}\beta^{-3/2},
                 \label{rrho_rot} \\
  Z^\#_{vib}&=& \prod_i\frac{1}{2\sinh(\omega_i\beta/2)},
                \label{rrho_vib}
\end{eqnarray}
where $\beta=1/k_BT$, $I_i$ are the principal inertia moments at the
saddle point, and $\omega_i$ are the frequencies for motions
transverse to the reaction coordinate.

Ad hoc approaches for accounting for the effects of tunneling often
assume a separability between the effects of tunneling and the
contribution of the transverse modes to the transition state partition
function.  Commonly, this assumption is implemented in terms of a
separability of the reaction coordinate and the transverse modes. Then
an additional tunneling factor enters into Eq.~\ref{rrho_notun},
\begin{eqnarray}
  Z^\#       &=&Z^\#_{tun}Z^\#_{rot}Z^\#_{vib},
                 \label{rrho} \\
  Z^\#_{tun}&=&\beta\int dE P(E) e^{-\beta E},
                \label{one_tun}
\end{eqnarray}
where $P(E)$ is the one-dimensional barrier transmission coefficient
at energy $E$. In the spirit of the RRHO approximation, a parabolic
barrier approximation can be used, for which the tunneling
transmission coefficent reads,
\begin{equation}
  \label{pb_tc}
  P(E) = \frac{1}{1 + e^{-2\pi E/\omega_b}},
\end{equation}
where $\omega_b$ is the imaginary frequency at the saddle point and
the saddle point is taken as the zero of energy. Substituting
Eq.~\ref{pb_tc} into Eq.~\ref{one_tun} one obtains the classic Wigner
tunneling expression,\cite{wigner32}
\begin{equation}
  \label{pb_tun}
  Z^\#_{tun}=\frac{\omega_b\beta/2}{\sin(\omega_b\beta/2)}.
\end{equation}

While Eq.~\ref{pb_tun} takes into account local tunneling effects, it
fails at low temperatures because the integral in Eq.~\ref{one_tun} diverges at low
energies for the parabolic barrier. This divergence occurs at temperatures lower than
the crossover temperature, $T_c$,
\begin{equation}
  \label{cross}
  k_BT_c= \frac{\omega_b}{2\pi}.
\end{equation}
To avoid this divergence one should use a more realistic model for the
transmission coefficient. In the one-dimensional WKB approximation,
the transmission coefficient reads,
\begin{equation}
  \label{wkb_tc}
  P(E) = e^{-\tilde{S}(-E)},\ E < 0,
\end{equation}
where $\tilde{S}(E)$ is the abbreviated action over the periodic trajectory along
the reaction coordinate $r$ with the energy $E$ in the inverted barrier
potential $-V(r)$,
\begin{equation}
  \label{one_act}
  \tilde{S}(E) = 2 \int_{r_-}^{r_+}dr \sqrt{2m_r(E+V(r))},
\end{equation}
where $r_\pm$ are the left and right turning points and $m_r$ is the
mass associated with the reaction coordinate. Evaluating the integral in
Eq.~\ref{one_tun} in the stationary phase approximation (SPA) one obtains,
\begin{equation}
  \label{one_deep}
  Z^\#_{tun}= \sqrt{2\pi} \beta\left(\frac{d^2\tilde{S}(E)}{dE^2}\right)^{-1/2}e^{-\tilde{S}(E) + \beta E},\
  \frac{d\tilde{S}}{dE} = \beta,
\end{equation}
which is valid in the "deep tunneling" regime where the Wigner
expression fails.

From Eqs.~\ref{one_act} and \ref{one_deep} one can see that, in
contrast with the localized TS dividing surface of relevance at higher
temperatures, in the deep tunneling regime the contribution from
tunneling through the TS is related to the dynamics over a relatively
wide range of coordinates. As a result, in the deep tunneling regime,
the separability assumption between the reaction coordinate and the
transverse modes becomes questionable and it is desirable to construct
a theory that avoids this assumption. Over the years, tremendous
effort has been devoted to constructing such a theory.  For
conciseness, we review only a few results of particular relevance to
our further discussion.

Long ago Langer\cite{langer67} suggested that the tunneling rate
constant can be calculated via analytical continuation of the system
partition function $Z= Z_R +\frac{i}{2}Z_{I}$, considered as a path
integral in the configurational space of the closed paths. The
imaginary part in this integral comes from the vicinity of the
trajectory that corresponds to the saddle point in that space and the
path integral is considered in the quadratic approximation. The TS
partition function $Z^\#$ in Eq.~\ref{tst} then reads,
\begin{equation}
  \label{lang}
  Z^\# = 2\pi Z_{I},
\end{equation}
Affleck\cite{affleck81} considered the transition from low to high
temperatures for one-dimensional systems and has confirmed that
Eq.~\ref{lang} coincides with Eq.~\ref{one_deep} at temperatures below
the crossover.

Many prior studies focused on the evaluation of the microcanonical
rate constant rather than the canonical one.  Miller\cite{miller75},
who was one of the pioneers in the study of muti-dimensional tunneling
effects in chemical reactions, applied his correlation function
formalism\cite{miller74} and Gutzwiller's expression for the Green's
function\cite{gutzwiller71} to obtain the expression for the
``cumulative reaction probability'', which we prefer to call the
microcanonical number of states, $N(E)$, in the spirit of RRKM
theory.\cite{marcus52} Notably, it was later realized that the
expression obtained for $N(E)$ did not agree with the exact expression
in the separable multi-dimensional case and a correction was
suggested.\cite{chapman75} Curiosly enough, Miller in his original
paper\cite{miller75} also suggested an expression for the thermal rate
constant that is correct in the separable case but his derivation was
inaccurate.\cite{comment1} Furthermore, as we see, this expression
appears to be missing key terms of relevance to non-separable
systems. Alternative, non-separable tunneling path approximations have
been developed for the ground vibrational state in the framework of
the vibrational adiabaticity assumption, starting with the seminal
paper of Marcus and Coltrin.\cite{marcus77} Truhlar and co-wokers, who
have originally developed many of these
approaches,\cite{scodje81,scodje82} extensively used them in thermal
rate constant
calculations.\cite{liu93,liu93a,ramos05,paneda10,paneda11,gao18}

Benderskii et al.\cite{benderskii92,benderskii93} further discussed
the derivation of the rate constant within Langer's approach.  By
neglecting the interaction between longitudinal and transverse
fluctuations, they obtained an explicit expression for the rate
constant that coincided with Miller's expression for the thermal
rate constant.\cite{miller75} It includes the product
$1/\prod_i2\sinh(u_i/2)$ as a prefactor, which looks formally similar
to the vibrational partition function, cf.
Eq.~\ref{rrho_vib}. Furthermore, the stability parameters $u_i$, which
were first introduced by Gutzwiller,\cite{miller75,gutzwiller71}
approach their classical values $u_i=\beta\omega_i$ when the
temperature approaches $T_c$, making the analogy with the vibrational
partition function complete. Jackels et al.\cite{jackels95}
contributed to the practical aspects of the tunneling path
calculations through a potential expansion in the vicinity of the
minimum energy path (MEP).
Richardson\cite{richardson16,richardson16a} has used an improved
Green's function approximation within Miller's correlation function
expressions.\cite{miller83} Many papers have successfully used the so
called ring polymer molecular dynamics approach for the tunneling
rate constant calculation.\cite{beyer16, richardson18} This approach
implements Langer's theory directly by approximating the partition
function path integral with a large number of beads.

To summarize, two generic classes of results are available in the
literature. These two classes differ in the presumed starting point
for the derivation. The papers based on Miller's correlation function
formalism express the rate constant in terms of a Boltzmann weighted
integral over energy. The integrand is expressed via Green's
functions, for which different approximations are used. The advantage
of this approach is that it is better justified and allows for a
smooth transition from low to high temperatures. Its drawback is that
current implementations do not reproduce the exact separable result,
cf. Eq.\ref{rrho}, because of the difficultities in evaluating the
Green's function.

On the other hand, the papers based on Langer's approach typically
either use different kinds of decoupling approximations to obtain the
final rate constant or calculate the fluctuation part of the path
integral directly. We did not find in the literature a closed rate
expression for the non-separable reactive system. So the main goal of
this paper is to provide such an expression that can be substituted
for Eq.~\ref{rrho} in the deep tunneling regime. Also, all analytical
papers we have seen, with one exception, do not take into account the
conservation of angular momentum. The one exception is a paper by
Miller\cite{miller75}, which, as we will discuss later, is
flawed. Therefore, the second goal of this paper is to fill this gap
by providing an expression for the analog of the rotational partition
function in the deep tunneling regime, and which goes smoothly to the
classical rotational partition function at higher temperatures. We
will also work through various ramifications of our results, focusing
on perturbative expansions.

The layout of the paper is as follows. In Sec.~2 we will discuss the
general features of the problem and consider the one-dimensional rate
expression. In Sec.~3 we will consider the coupled reaction
coordinate + vibrational modes problem when the energy is the only
integral of motion. The conservation of angular momentum is
considered in Sec.~4. The paper concludes with a lengthy general
discussion, Sec.~5 and some brief summarizing remarks, Sec. 6.

\setcounter{equation}{0}
\section{General Formalism and One-Dimensional Case}
We will use Langer's approach\cite{langer67} to calculate the
TS partition function at low temperatures.  Let us
consider the Hamiltonian system with $N$ degrees of freedom with
unit masses in the potential $\bar{V}(q)$,
\begin{equation}
  \label{main_ham}
  \hat{H} = -\frac{1}{2}\sum_i\frac{\partial^2}{\partial q_i^2}+\bar{V}(q).
\end{equation}
In what follows we will often use vector notation. For example,
$\dot{q}^2$ will mean $\sum_i\dot{q}_i^2$.

The density matrix $\hat{\rho} = e^{-\beta\hat{H}}$ can be written in
the form of the path integral,\cite{feynman72}
\begin{eqnarray}
  \label{dm_pi}
  \rho(q_1, q_2; \beta) &=& \pint q\exp(-S[q]),
  \\
  \label{tun_act}
  S[q]  &=& \int_0^T dt(\dot{q}^2/2 - V(q)),
  \\
  \nonumber
  T&=&\beta,\ q(0)=q_1,\ q(T)=q_2.
\end{eqnarray}
The function $S[q]$ formally coincides with the action of the
classical particle in the inverted potential $V(q)=-\bar{V}(q)$ and
$T$ denotes the trajectory period in this section (not to be confused
with the temperature).

The partition function of the system can be written as
$Z=Tr[\hat{\rho}]=\int d^N\! q\rho(q,q;T)$.  The tunneling trajectory,
also known as the instanton trajectory, $q_0(t)$ corresponds to the
saddle point of the action $S$ in the configurational space of the
closed paths.  The instanton trajectory is the key object in Langer's
theory. Considering small variations around the instanton path, one
comes to the conclusion that it satisfies the classical equations of
motion,
\begin{equation}
  \label{ins_em}
  \ddot{q}_0 =-\frac{\partial V}{\partial q}_0,\  S(q_1,q_2,t)=S[q_0].
\end{equation}
Then, taking into account that
\begin{equation}
  \label{act_rel}
  \dot{q}_2=\frac{\partial S(q_1,q_2, t)}{\partial q_2},\ \
  \dot{q}_1 = -\frac{\partial  S(q_1, q_2, t)}{\partial q_1},
\end{equation}
and that the first derivative of $S$ over $q_1=q_0(0)=q_2=q_0(T)$ should be zero,
one finds that the initial and final momenta coincide,
$\dot{q}_0(0)=\dot{q}_0(T)$, that is the instanton trajectory is
periodic. In the case of a separable reaction coordinate the
instanton trajectory coincides with that described in the introduction for the
WKB approximation.

The tunneling action in the quadratic approximation about the
instanton trajectory reads,
\begin{equation}
  \label{ins_act}
  S'[\delta q] = \frac{1}{2}\int_0^T dt (\delta \dot{q}^2 -
  \sum_{i,j}\frac{\partial^2V(q_0(t))}{\partial q_i\partial
    q_j}\delta q_i \delta q_j).
\end{equation}
In the quadratic approximation the partition function $Z_{I}$ can be
written as
\begin{eqnarray}
  \label{ins_pf}
  &&iZ_{I} = \exp(-S(T))\int d^N\! q'\pint q \exp(-S'[q]),
  \\
  \nonumber
  &&q(0)=q(T)=q',
\end{eqnarray}
where $S(T)=S[q_0]$. The expression under the integral can be
viewed as the infinite dimensional gaussian integral,
\begin{equation}
  \label{gauss}
  Z_{I}\sim\int dq_tdq_2...exp(-\lambda_1q_1^2  -\lambda_2q_2^2 - ...).
\end{equation}
One $\lambda_i$ in this equation is negative. It corresponds to
the unstable mode at the saddle point. After analytical continuation
it should give a purely imaginary value. We will see that it happens
automatically when we calculate the path integral in
Eq.~\ref{ins_pf}.

There is another $\lambda_i$ that is identically zero. This feature is
related to the fact that the Lagrangian equations, Eq.~\ref{ins_em},
are invariant under a time shift and therefore the instanton
trajectory is not a point in the configurational space of the closed
paths but rather a circle, which is obtained by an arbitrary time
shift transformation to the individual instanton trajectory. The
linearized trajectory, which corresponds to the infinitesimal
time-shift and which has zero $\lambda_i$, is the derivative of the
original instanton $q_0(t)$ over the shift $\tau$, $\dot{q}_0(t)$. The
way to handle this divergence is to leave the corresponding gaussian
integral,
\begin{equation}
  \label{zero_int}
\int\exp(-S'[\dot{q}_0]x^2)dx,
\end{equation}
in Eq.~\ref{ins_pf} intact and substitute it with the time shift
transformation period, which is $T=\beta$.

Actually, any continuously parametrized transformation that preserves
the Lagrangian in Eq.~\ref{ins_act} will generate an additional
dimension for the instanton manifold and a corrseponding $\lambda_i$
with zero value. So, for the angular momentum conservation case one
will have four zero $\lambda_i$'s.  To handle the three additional
divergent gaussian integrals in Eq.~\ref{ins_pf} one again leaves
these integrals, which correspond to three infinitesimal orthogonal
rotations, intact and substitutes them with the total solid volume of
the three dimensional rotational group, which equals to $8\pi^2$.

To proceed we use Feynman's result about path integrals with
quadratic action\cite{feynman72}. He showed that the path integral
$K(q_2,q_1,t)$
with the action $S'[q]$ given by  Eq.~\ref{ins_act} for trajectories,
which start at time
$t=0$ with $q=q_1$ and finish at time $t$ with $q=q_2$ can be written as 
\begin{eqnarray}
  \label{qpi}
  &&K(q_2,q_1,t)= U(t)e^{-S'(q_1,q_2,t)},
  \\
  \nonumber
  &&S'(q_1,q_2,t) =S'[\bar{q}], 
\end{eqnarray}
where $\bar{q}$ is the trajectory that satisfies the linearized
equations of motion,
\begin{equation}
  \label{ins_lem}
  \ddot{\bar{q}}_i = -\sum_j\frac{\partial^2V(q_0(t))}{\partial q_i\partial q_j}\bar{q}_j.
\end{equation}
with $\bar{q}(0)=q_1$ and $\bar{q}(t)=q_2$. Further we will consider
only the trajectories which satisfy Eq.~\ref{ins_lem} and will omit
the bar over $q$ notation. The function $U(t)$ does not depend on
$q_1$ and $q_2$. In Appendix A we show that $U(t)$ is expressed in
terms of the final coordinates of $N$ trajectories
$q^{(1)}(t), q^{(2)}(t),\ ...\ q^{(N)}(t)$, which start with the
initial velocities
$\dot{q}^{(1)}(0), \dot{q}^{(2)}(0),\ ...\ \dot{q}^{(N)}(0)$, and zero
initial coordinates as (see also Ref.\cite{althorpe11}),
\begin{equation}
  \label{pref}
  U(t) =(2\pi)^{-N/2}\sqrt{\frac{|\dot{q}^{(1)}(0),\dot{q}^{(2)}(0),...,\dot{q}^{(N)}(0)|}{|q^{(1)}(t),q^{(2)}(t),...,q^{(N)}(t)|}}.
\end{equation}

Substituting Eqs.~\ref{qpi} and \ref{pref} into Eq.~\ref{ins_pf} one
obtains the final expression for $Z_{I}$,
\begin{equation}
  \label{ins_pf_res}
  iZ_{I}=e^{-S(T)}U(T) \int' d^N\! q' e^{-S'(q')},
\end{equation}
where $S'(q') = S'(q',q',T)$.  The prime notation on the integral
means that the divergent terms must be substituted with the
appropriate values as described above.  It is easy to see that the
following expression for the tunneling action $S'(q')$ holds,
\begin{equation}
  \label{act_rel1}
  S'(q') = \frac{1}{2}q'(\dot{q}(T)-\dot{q}(0)).
\end{equation}
Thus, all necessary quantities, cf. Eqs.~\ref{pref} and
\ref{act_rel1}, are expressed via the initial and final values of the
coordinates and momenta. In what follows we will refer to them at
$t=0$ unless it is explicitly stated otherwise.

Using the obtained relations let us consider the one-dimensional case.
We will choose the starting point of the instanton trajectory away
from the turning point, $\dot{q}_0 \neq 0$.  To calculate $S'(q)$,
Eq.~\ref{act_rel1}, we need to find the trajectory for which
$q(T) = q$. But we know such a trajectory. It is $\dot{q}_0(t)$. We
note that up to the $\dot{q}_0$ factor the corresponding integral
coincides with Eq~\ref{zero_int}.

To caculate $U(T)$ we need to find another trajectory that
satisfies Eq.~\ref{ins_lem}. There are different ways to do so.
We choose a way that proves useful later for the multi-dimensional case.
Let us consider $q_0$ as a function of energy. Then the trajectory
\begin{equation}
  \label{en_traj}
  q_E(t)=\frac{\partial q_0(t, E)}{\partial E}
\end{equation}
satisfies the linearized equations of motions, Eq.~\ref{ins_lem},
because $q_0(t, E)$ satisfies Eq.~\ref{ins_em} and it is linearly
independent of $\dot{q}_0$. The latter can be seen from the fact that
the linearized energy, which is an integral of motion for the
linearized equations of motion, Eq.~\ref{ins_lem}, is zero for the
$\dot{q}_0$ trajectory and unity for the $q_E$ one.  Without loss of
generality one can assume that $q_E(0) = 0$. Then, from the energy
expression $E = \dot{q}^2/2 + V(q)$ one obtains
\begin{equation}
  \label{qe_beg}
  \dot{q}^E= 1/\dot{q}_0.
\end{equation}
To obtain the value of $q_E(t)$ at $t=T$ one notes that
$q_0(t, E + \delta E)$ is also a periodic trajectory with a slightly
different period $T(E + \delta E) = T + \frac{dT}{dE}\delta E$.  Then
it is easy to see that
\begin{equation}
  \label{qe_end}
  q_E(T) = - \frac{dT}{dE}\dot{q}_0. 
\end{equation}
Note the sign in this equation. This means that the tunneling trajectory $q_0(t)$
has passed through the focal point and the determinant ratio in
Eq.~\ref{pref} has a negative sign. Substituting Eqs.~\ref{qe_beg} and
\ref{qe_end} into Eq.~\ref{pref} one obtains
\begin{equation}
  \label{u_exp}
  U(T) = \sqrt{-1}\left(\dot{q}_0\sqrt{2\pi\frac{dT}{dE}}\right)^{-1}.
\end{equation}
Substituting Eq.~\ref{u_exp} and $\beta$ for
$\int\exp[-S'(\dot{q}_0)x^2]dx$ into Eq.~\ref{ins_pf} and then into
Eq.~\ref{lang} one obtains,
\begin{equation}
  \label{one_ins}
  Z_{tun}=\beta \sqrt{2\pi}\left(\frac{dT}{dE}\right)^{-1/2}e^{-S(T)}.
\end{equation}
To match with Eq.~\ref{one_deep}, one notes that the full action $S(q',
q'', t)$,
Eq.~\ref{tun_act}, considered as a function of time, is related with
the abbreviated action $\tilde{S}(q', q'', E)$,
\begin{equation}
  \label{short_act}
  \tilde{S}(q', q'', E) = \int \dot{q}dq,
\end{equation}
considered as a function of energy, by Legendre's transform,
\begin{equation}
  \label{act_leg}
  \tilde{S} = S + Et,\ \frac{\partial S}{\partial t} = -E,\ \frac{\partial\tilde{S}}{\partial E} = t.
\end{equation}
Then, taking into account Eq.~\ref{act_rel} and that the tunneling
trajectory is periodic, one obtains that $T=\frac{d\tilde{S}}{dE}$ and
Eq.~\ref{one_deep} is recovered.

\setcounter{equation}{0}
\section{Reaction Coordinate + Vibrational Modes Case}
In this section we will denote $\dot{q}$ as $p$ and the state vector
$(q, p)$ as $v$.  Let us consider the linear $2N$-dimensional
transformation $\hat{M}$, known as the monodromy matrix,\cite{gutzwiller71}
from $v(0)$ to $v(T)$
\begin{equation}
  \label{mon}
  v(T)=\hat{M}v(0).
\end{equation}
For Hamiltonian systems the monodromy transformation is symplectic, as
it preserves the natural symplectic form, $\hat{\Omega} = q\wedge p$,\cite{arnold78}
\begin{equation}
  \label{symp}
  \hat{\Omega}(v_1, v_2) = q_1p_2 - q_2p_1.
\end{equation}
We will call the operation $\hat{\Omega}(v_1, v_2)$ the symplectic product of
vectors $v_1$ and $v_2$ and denote it as $v_1\odot v_2$.

The eigenvalues of a symplectic transformation come in
pairs, $\lambda_+\lambda_-=1$. \cite{arnold78} To each
$\lambda_i \ne 1$ there corresponds a unique eigenvector
$v_i,\ \hat{M}v_i=\lambda_iv_i$. For the unit eigenvalue the situation is
more complicated and the corresponding eigenspace  is transformed
through itself.

Now we turn to the analysis of the monodromy transformation for our
problem. The eigenvectors with $\lambda \ne 1$ correspond to the
vibrational transverse modes. With energy being the only integral of
motion, there are $2(N - 1)$ of them.  We will denote them as
$v_{i,\pm}=(q_{i,\pm}, p_{i,\pm}),\ i=1,..,N-1$.  We will also denote
$\lambda_{i,+}$ as $\lambda_i$.  To proceed we will assume that the
tunneling trajectory has a turning point where the system comes to
complete rest, $\dot{q}_0=0$, and, for a moment, that the $t=0$ moment
corresponds to it. Then, due to time-reversal symmetry, the vectors
$v_{i,+}$ and $v_{i,-}$ correspond to the trajectories that are
time-reversed relative to each other, so that $q_{i,+} = q_{i,-}=q_i$
and $p_{i,+}=-p_{i,-}=p_i$. Considering the symplectic product
$v_{i,s}\odot v_{j,r}$ before and after the monodromy transformation
for each pair of $v_{i,s}$ and $v_{j,r}$ one finds that
\begin{equation}
  \label{ort_rel0}
  p_iq_j=\delta_{i,j}
\end{equation}
with a proper normalization. If $p_iq_i<0$ one can always exchange
$v_{i,+}$ and $v_{i,-}$.

The two lineary independent vectors which are left are associated with
the reaction coordinate. They correspond to the $\dot{q}_0(t)$ and
$q_E(t)=\frac{\partial q_0(t, E)}{\partial E}$ linearized trajectories
and constitute the eigenspace for $\lambda = 1$. Using the same
arguments that have been used to derive Eq.~\ref{qe_end}, one 
obtains
\begin{equation}
  \label{qe_mon}
  \hat{M}v_E=v_E-\frac{dT}{dE}v_0.
\end{equation}
Using Eq.~\ref{symp} and the linearized energy form,
\begin{equation}
  \label{en_int}
  H(q, p) = g_0q,\ \ g_0=\frac{\partial V}{\partial q}_0,
\end{equation}
($g_0$ is the potential gradient at the turning point)
the following additional orthogonality relations can be
obtained,
\begin{eqnarray}
  \label{ort_rel1}
  v_0\odot v_i&=& g_0q_i=0,
  \\
  \label{ort_rel2}
  v_E\odot v_i&=&q_Ep_i = 0,
  \\
  \label{ort_rel3}
  q_Eg_0&=& 1.
\end{eqnarray}

We will proceed with calculation of the prefactor $U(T)$.  One can
use $p_i$ as starting momenta in Eq.~\ref{pref} for the unstable
trajectories.  Then the corresponding coordinates, obtained after
$t=T$, are given by
\begin{equation}
  \label{qf_exp}
q^f_i=\frac{1}{2}(\lambda_i - \lambda_i^{-1})q_i.
\end{equation}

Unfortunately, one cannot use the turning point $q_\pm$ as the starting
point 
for one of the reaction coordinate trajectories $\dot{q}_0(t)$ and
$q_E(t)$, because $q_\pm$ is a focal point for the periodic trajectory
that has been started from it, cf. Eq.~\ref{u_exp}. Therefore we will
make a small time-shift $\tau$ from the turning point and use 
perturbation theory to calculate the determinants ratio in
Eq.~\ref{pref}.  The calculation is rather cumbersome and therefore
has been moved to Appendix B. The result is shown below,
\begin{eqnarray}
  \label{u_exp1}
  U(T)&=& (2\pi)^{-N/2}\tau^{-1}
          \prod_i\left[\frac{1}{2}(\lambda_i-\lambda_i^{-1})\right]^{-1/2}\sqrt{-\frac{|g_0,p_i|}{|g_0,q_i|}}
  \\
  \nonumber
  &\times&\left(\frac{dT}{dE}g_0^2+\frac{2}{g_0^2}\sum_i (p_ig_0)^2
    \frac{\lambda_i+\lambda_i^{-1}-2}{\lambda_i-\lambda_i^{-1}}\right)^{-1/2}.
\end{eqnarray}
Here and in what follows we use the shorthand notation for the group of
vectors under the determinant, for example,
$|g_0,p_i|=|g_0,p_1,p_2,...|$.  It is worthwhile noting again the
minus sign under the square root in this equation.

Now we turn to calculating the action integral in
Eq.~\ref{ins_pf_res}. For the action one can use Eq.~\ref{act_rel1}.
Due to Eq.~\ref{ort_rel0} the action is diagonalized if one uses $q_i$
as the new basis. Expressing the linearized trajectory $v(t)$, for which
$q(0)=q(T)=q_i$, in terms of the linear combination of the
eigenvectors, $v(t) = C_{i,+}v_{i,+}(t)+C_{i,-}v_{i, -}(t)$, at $t=0$ and
$t=T$, one obtains
$C_{i,+} + C_{i,-} = \lambda_iC_{i,+}+\lambda_i^{-1}C_{i,-}= 1$, which
gives $C_{i,+}=1/(1+\lambda_i),\;C_{i,-}= \lambda_i/(1+\lambda_i)$.
Substituting $C_{i,\pm}$ into Eq.~\ref{act_rel1},
$\dot{q}(T)-\dot{q}(0)=(C_{i,+}\lambda_i-C_{i,-}/\lambda_i-C_{i,+}+C_{i,-})p_i$,
the expression for the action coefficient of the diagonal term reads
$(\lambda_i - 1)/(\lambda_i + 1)$, where we again have used
Eq.~\ref{ort_rel0}.

The zero-mode term in the integral in Eq.~\ref{ins_pf_res} is reduced
to the standard form, Eq.~\ref{zero_int}, if one uses
$\dot{q}_0=-g_0\tau$ as the basis vector in it, which corresponds to
the $\dot{q}_0(t)$ trajectory.  Substituting the zero-mode standard
integral with $\beta$, and using Eq.~\ref{u_exp1} one obtains the
final expression for $Z^\#$,
\begin{eqnarray}
  \label{multi_ins}
  Z^\#&=&\beta\sqrt{2\pi}F_{tun}\prod_i[2\sinh(u_i/2)]^{-1}e^{-S(\beta)},
  \\
  \label{tun_pref}
  F_{tun}&=&\left(\frac{d^2\tilde{S}}{dE^2}+\frac{2}{g_0^4}\sum_i
             (p_ig_0)^2\tanh(u_i/2)\right)^{-1/2},
  \\
  \label{s0}
  S(\beta)&=&\tilde{S}(E)-E\beta,\ \frac{d\tilde{S}}{dE}=\beta,
\end{eqnarray}
where we have used the conventional stability parameters $u_i$,
$\lambda_i=e^{u_i}$.  One can see that the tunnneling prefactor term given
by Eq.~\ref{tun_pref} differs from the corresponding one-dimensional
one $d^2\tilde{S}/dE^2$, cf Eq.~\ref{one_ins}, by the presence of an
additional term.

\setcounter{equation}{0}
\section{Angular Momentum Conservation Case}
In the case when the potential is invariant under spatial rotations
there are three additional integrals of motion for Eq.~\ref{ins_em},
which are the angular momentum $J=q\times \dot{q}$ components. We will
use the vector notation $a\times q$, which could mean a vector product
$a\times q_i$ for each $i$-th atom or the sum
$\sum_{i=1}^{N_a}a_i\times q_i$ depending on the nature of $a$.

We will assume that $t=0$ corresponds to the turning point. Then, in
the linearized form, the angular momentum integral of motion is given
by
\begin{eqnarray}
  \label{am_int}
  J(q, p) = q_0\times p.
\end{eqnarray}
Together with Eq.~\ref{en_int} they can be used to classify linearly
independent trajectories.

There are $2(N - 4)$ unstable modes ($N=3N_a-3$ is the number of
vibrations and rotations), which correspond to the transverse
vibrations. As in the previous section, $v_{i,\pm}= (q_i,\pm p_i)$
with the eigenvalues ($\hat{M}v_{i,\pm}=\lambda_{i,\pm}v_{i,\pm}$)
$\lambda_{i,\pm}=\lambda_i^{\pm1}$. The vectors $p_i$ and $q_i$
satisfy the orthogonality relations, Eq.~\ref{ort_rel0}.

The two vectors associated with the reaction coordinate are
$v_0=(0, -g_0)$ and $v_E=(q_E, 0)$. Together with $v_i$ they satisfy
Eqs.~\ref{ort_rel1}-\ref{ort_rel3}. While $v_0$ is the eigenvector of
the monodromy transformation $\hat{M}v_0=v_0$ with unit eigenvalue,
$v_E$ satisfies Eq.~\ref{qe_mon}.  It is worth noting that, due
to rotational invariance of the potential, the turning point actually
is not a point anymore but rather a three-dimensional sphere in the
$q$-subspace. As a result, the vector $q_E$ is defined up to
an arbitrary linearized rotation $n\times q_0$. We will choose it
so that
\begin{equation}
  \label{ort_rel9}
  q_E\times q_0=0.
\end{equation}

There are six additional vectors associated with the angular
momentum. Three of them are obtained by infinitesimal rotations of the
instanton trajectory,
\begin{equation}
  \label{am_rot}
v_\xi=(q_\xi,0),\  q_\xi=n_\xi\times q_0,\ \xi=x,y,z,
\end{equation}
whery $n_\xi$ is the unit vector in the direction of $\xi$. Because
the potential is invariant relative to rotations, vectors $q_\xi$
satisfy the relations,
\begin{equation}
  \label{ort_rel4}
  q_\xi g_0=0.
\end{equation}
Vectors $v_\xi$ are the eigenvectors of the monodromy transformation
$\hat{M}v_\xi=v_\xi$ with unit eigenvalue. Applying the monodromy
transformation to $v_\xi\odot v_{i,\pm}$ one obtains,
\begin{equation}
  \label{ort_rel5}
  v_\xi\odot v_{i,\pm}= q_\xi p_i =0,\ q_0\times p_i=0.
\end{equation}

Three other vectors $v^J_\xi,\ \xi=x,y,z$ correspond to non-zero
values of the angular momentum, Eq.~\ref{am_int}. We will choose them
in the following way.  In the $q$-space the vectors $q_E$, $q_\xi$,
and $q_i$ constitute the complete basis. The vectors $g_0$ and $p_i$
are complemetary to $q_E$, $q_\xi$, and $q_i$,
cf. Eqs.~\ref{ort_rel0}, \ref{ort_rel1}-\ref{ort_rel3},
\ref{ort_rel4}, and \ref{ort_rel5}. But they are incomplete. We will
add to them the vectors $p_\xi,\ \xi=x,y,z$ so that they will form a 
complete basis that is complementary to the basis $q_E$, $q_\xi$,
and $q_i$,
\begin{eqnarray}
  \label{ort_rel6}
  p_\xi q_\eta&=&\delta_{\xi,\eta},
  \\
  \label{ort_rel7}
  p_\xi q_i    &=& 0,
  \\
  \label{ort_rel8}
  p_\xi q_E   &=& 0.
\end{eqnarray}
Then the vectors $v^J_\xi$ can be defined as,
\begin{equation}
  \label{am_dyn}
  v_\xi^J=(0, p_\xi),\ \xi=x,y,z.
\end{equation}
Substituting $v_\xi$ into Eq.~\ref{am_int} one finds that
$Jv_\xi=n_\xi$.  The symplectic product $v_\xi\odot v^J_\eta$ is given
by,
\begin{equation}
  \label{sym_rel1}
v_\xi\odot v^J_\eta=\delta_{\xi,\eta}.  
\end{equation}
and the following products are zeros,
\begin{equation}
  \label{sym_rel2}
v_E\odot v^J_\xi=0,\ v_i\odot v^J_\xi=0,
\end{equation}
cf. Eqs.~\ref{ort_rel6}-\ref{ort_rel8}.

The result of the application of the monodromy matrix to
$v^J_\xi$ can be written in its most general form as,
\begin{equation}
  \label{jv_mon}
  \hat{M}v_\xi^J= v_\xi^J + C_0^\xi v_0 + \sum_{i,s} C_{i,s}^\xi v_{i,s}
  +\sum_{\eta} C_{\xi,\eta} v_{\eta},
\end{equation}
where we have used the integrals of motion, Eqs.~\ref{en_int} and
\ref{am_int}, to exclude $v_E$ and $v_\eta^J,\ \eta\ne\xi$ terms.
Considering the symplectic products $v_E\odot v^J_\xi$ before and
after the monodromy transformation one concludes that
$C_0^\xi=0$. Similarly, considering the symplectic products
$v_{i,\pm}\odot v^J_\xi$, one finds that $C_{i,\pm}^\xi=0$.  As a
result, Eq.~\ref{jv_mon} is considerably simplified,
\begin{equation}
  \label{jv_mon1}
  \hat{M}v_\xi^J= v_\xi^J  +\sum_{\eta} C_{\xi,\eta} v_{\eta}. 
\end{equation}
Considering the symplectic products
$v_\xi^J\odot{v}^J_\eta$ before and after the monodromy
transformation and using Eq.~\ref{sym_rel1}
one finds that the matrix $C_{\xi,\eta}$ is symmetric,
\begin{equation}
  \label{sym_rel3}
  C_{\xi,\eta}=C_{\eta,\xi}.
\end{equation}
From Eq.~\ref{jv_mon1} it follows that if the matrix $C_{\xi,\eta}$
is not degenerate, the six-dimensional unit eigenspace related to the
angular momentum is factorized into three two-dimensional subspaces
$\tilde{v}_\xi^J,\ \sum_{\eta} C_{\xi,\eta} v_{\eta},\ \xi=x,y,z$.

Now we turn to the calculation of the prefactor $U(T)$,
Eq.~\ref{pref}. For the same reason as in the previous section we
apply to the system a small time-shift and consider the corresponding
trajectories using perturbation theory.  The corresponding calculation
is pretty much the same as in Appendix B for the system without
angular momentum conservation. The differences are traced in Appendix
C. Thus, the expression for $U(T)$ is given by, cf. Eq.~\ref{c19},
\begin{eqnarray}
 \label{u_exp_am}
  U(T)&=& (2\pi)^{-N/2}\tau^{-1}
             \sqrt{-\frac{|g_0,p_i,p_\xi|}{|g_0,q_i,\sum_{\eta}C_{\xi,\eta}q_{\eta}|}}
             \prod_i\left[\frac{1}{2}(\lambda_i-\lambda_i^{-1})\right]^{-1/2}
  \\
  \nonumber
      &\times&\left(\frac{dT}{dE}g_0^2+\frac{2}{g_0^2}\sum_i (p_ig_0)^2
               \frac{\lambda_i+\lambda_i^{-1}-2}{\lambda_i-\lambda_i^{-1}}\right)^{-1/2}
\end{eqnarray}
where $C_{\xi,\eta}$ is the matrix from Eq.~\ref{jv_mon1}. One can
simplify this expression by multiplying the numerator and denominator
under the square root with $|g_0q_iq_\xi|$ and then consider the
matrix product $(g_0q_iq_\xi)^T\times(g_0p_ip_\xi)$, whose determinant
gives $g_0^2$,
\begin{eqnarray}
  \label{u_exp_am1}
  U(T)&=& \sqrt{-1}(2\pi)^{-N/2}\tau^{-1}|g_0q_iq_{\xi}|^{-1}|C|^{-1/2}
  \prod_i\left[\frac{1}{2}(\lambda_i-\lambda_i^{-1})\right]^{-1/2}.
  \\
  \nonumber
      &\times&\left(\frac{dT}{dE}+\frac{2}{g_0^4}\sum_i (p_ig_0)^2
               \frac{\lambda_i+\lambda_i^{-1}-2}{\lambda_i-\lambda_i^{-1}}\right)^{-1/2}
\end{eqnarray}

The integration of the exponent in Eq.~\ref{ins_pf_res} is performed in a
similar manner as in the previous section. However, to handle the
divergence associated with the rotational invariance of the tunneling
trajectory, one has to recover the original integrals over the
trajectories corresponding to the infinitesimal orthogonal rotations
and then substitute them with the total solid angle,
as described in Sec.~2. This is achieved by using
$q_\xi,\ \xi=x,y,z$ as the basis vectors in the integral in
Eq.~\ref{ins_pf_res}. Collecting all the terms, one comes to the
following expression,
\begin{equation}
  \label{exp_int_am}
  \int' d^N\! q' e^{-S'(q')} = 2(2\pi)^{-N/2}\beta\tau|g_0q_iq_\xi|
  \prod_i\sqrt{\frac{1}{2}\frac{\lambda_i+1}{\lambda_i-1}},
\end{equation}
where the prime index on the integral indicates that the
divergences have been replaced by the proper values.

Substituting Eqs.~\ref{u_exp_am1} and \ref{exp_int_am} into
Eq.~\ref{ins_pf_res} and then into Eq.~\ref{lang}, one obtains for
$Z^\#$,
\begin{equation}
  \label{ins_rrho}
  Z^\#=4\pi{\beta}F_{tun}\prod_i[2\sinh(u_i/2)]^{-1}|C|^{-1/2}e^{-S(\beta)},
\end{equation}
where the tunneling prefactor $F_{tun}$ is given by Eq.~\ref{tun_pref}
and the matrix $C_{\xi,\eta}$ is given by Eq.~\ref{jv_mon1}.

\setcounter{equation}{0}
\section{Discussion}
\subsection{General Observations and Interpretation}
Eqs.~\ref{multi_ins}, \ref{tun_pref}, and \ref{ins_rrho} together with
Eq.~\ref{jv_mon1} are the main result of this paper. They provide a
closed expression for the tunneling rate constant in terms of the
properties of the tunneling trajectory.  In deriving them we have made
no assumptions about system separability.  Our expression for the
multi-dimensional term in the prefactor, Eq.~\ref{tun_pref}, differs
from the corresponding one-dimensional factor
$\frac{d^2\tilde{S}}{dE^2}$ by the additional term,
\begin{equation}
  \label{add_tun}
 F_{tun}'=\frac{2}{g_0^4}\sum_i (p_ig_0)^2\tanh(u_i/2),
\end{equation}
which is a direct consequence of the system non-separability.  This term
disappears for separable systems, because $g_0p_i$ is then $0$.  For
non-separable systems this term should be present (it disappears only
if all $p_ig_0=0$).  At small energies, when the temperature
approaches the crossover $T_c$, the direction of $g_0/|g_0|$ is close
to the direction of the unstable imaginary vibrational mode and
$g_0^2\propto E$. On the other hand, the projection of $p_i/|p_i|$ on
the direction of $g_0/|g_0|$ also tends to zero, because $p_i/|p_i|$
becomes close to the direction of the corresponding stable transverse
vibration mode. As a result, the expression in Eq.~\ref{add_tun}
contains the ratio of two small terms.

In Appendix D we show that for a cubic anharmonicity in the potential,
cf. Eq.~\ref{multi_pot_exp} below, when the temperature is close to
the crossover and, correspondingly, the energy of the instanton
trajectory is close to zero, $F_{tun}'$ is given by,
\begin{equation}
  \label{add_tun_cub}
  F_{tun}'=8\sum_iV^{(3)^2}_{0,0,i}\omega_i^{-3}(\omega_i^2+4\omega_b^2)^{-2}\tanh(u_i/2),\
  u_i=2\pi\frac{\omega_i}{\omega_b}.
\end{equation}
The importance of the $F_{tun}'$ term depends on its ratio to
$d^2\tilde{S}/dE^2$, cf. Eq.~\ref{tun_pref}.  In Table~\ref{tun_tab}
below the properly normalized term $\frac{1}{2}\omega_b^2F_{tun}'$ is
compared with
$\alpha=\frac{1}{2}\omega_b^2\left.\frac{d^2\tilde{S}}{dE^2}\right|_{E=0}$
for several common reactions.  It seems that in most situations,
unless $\alpha$ is anomalously small or even negative (see the
discussion below), the contribution of the $F_{tun}'$ term is small in
comparison with the main term.
    
In the framework of Langer's theory the prefactor to the tunneling
exponent in the expression for the rate constant is related to
fluctuations arround the instanton trajectory. In the separable case
the fluctuations associated with the energy change are related to the
reaction coordinate and untangled from the fluctuations with the same
energy. The first ones give the factor
$(\frac{d^2\tilde{S}}{dE^2})^{-1/2}$ in the rate constant expression
and the rest give the TS partition function in the harmonic
approximation. In the non-separable case all fluctuations are
entangled and this results in the appearence of the additional term in
Eq.~\ref{tun_pref}. Therefore, we will call the
$\frac{1}{2}\omega_b^2F_{tun}'$ term the entanglement factor (EF).

In the case when the angular momentum is conserved, an additional
factor appears in the rate expression, cf. Eqs.~\ref{multi_ins} and
\ref{ins_rrho}, which looks similar to the inertia moments matrix in
the classical rate constant expression, cf. Eq.~\ref{rrho_rot}. One
can show that when the temperature
approaches $T_c$ this factor is indeed transformed into
$\beta^{-3/2}\sqrt{I_xI_yI_z}$. To this end let us assume that the
system of coordinates is chosen to diagonalize the inertia moments
matrix and consider the following auxiliary vectors
\begin{equation}
  \label{aux_vec}
v^{aux}_\xi=(0, I_\xi^{-1}q_\xi),
\end{equation}
where $I_\xi$ are the principal inertia moments. Note the difference
with $v_\xi=(q_\xi,0)$. We will look for $v^{aux}$ in the form
$I_\xi^{-1}q_\xi=A_0g_0+\sum_{\eta}A_{\eta}p_\eta+\sum_iA_{\xi,i}p_i$.
Considering the scalar product with $q_E$ and taking into account
Eq.~\ref{ort_rel9}, one finds that $A_0=0$. Considering the scalar
product with $q_\zeta$ one finds that
$A_\eta=\delta_{\eta,\xi}$. Finally, considering the scalar product
with $q_i$ one finds that $A_{\xi,i}= I_\xi^{-1}q_\xi q_i$. Thus,
\begin{equation}
  \label{aux_vec1}
I_\xi^{-1}q_\xi=p_\xi + I_\xi^{-1}\sum_i(q_\xi q_i)p_i.
\end{equation}
The vector $v^{aux}_\xi$ corresponds to the pure rotation arround the
$\xi$-axis with unit angular momentum and the angular velocity of
$I_\xi^{-1}$. When the temperature approaches $T_c$ the tunneling
trajectory corresponds to a small vibration-like motion around the
saddle point. To a good approximation, in this limit the rotation
should be uncoupled from vibrations, $q_\xi q_i\simeq0$ and
$v^{aux}_\xi\simeq v^J_\xi$. Then, after ``time'' $\beta$ the system
will turn around the $\xi$ by the angle $\beta I_\xi^{-1}$. Thus,
$Mv^J_{\xi}\simeq v^J_\xi + \beta I_\xi^{-1}q_\xi$. Comparing this
equation with Eq.~\ref{jv_mon1} one concludes that
$C_{\xi,\eta}\simeq\beta I_\xi^{-1}\delta_{\xi,\eta}$. Substituting
this equation into Eq.~\ref{ins_rrho} one obtains the classical
rotation function contribution, Eq.~\ref{rrho_rot}.

Miller\cite{miller75} has suggested a different way to account for
angular momentum conservation, which is formally analogous to the way
that the energy is treated in his correlation function
approach. Namely, he suggested to calculate the corresponding Green's
functions $G(q, q, E, J)$ postulating and using periodic trajectories
that are functions of both $E$ and $J$. We believe that this approach is
incorrect because such trajectories do not exist.  The dependence on
$J$ is rather reminiscent of the dependence of the tunneling
trajectory on energy at high temperatures: it collapses to a point. So
for each energy there is only one periodic trajectory, for which
$J=0$. As an additional argument against considering the periodic
trajectories with non-zero angular momentum one may note that while
the trajectory in real time corresponds to real $J$, the trajectory in
imaginary time would correspond to pure imaginary $J$, in contrast
with the energy, for which the motion in imaginary time coresponds to
potential inversion. Thus, the trajectory with real $J$ in imaginary
time would be unavoidably complex. Perhaps these difficulties are why
subsequent studies do not appear to have followed up on this
suggestion.

Equation~\ref{ins_rrho} is an analog of the RRHO rate constant,
Eq.~\ref{rrho}, at low temperatures. For practical applications it is
convenient to formally partition it into two parts, the effective TS
partition function $Z^\#_{RV}$ and the effective tunneling factor
$Z^\#_{tun}$,
\begin{eqnarray}
  \label{rv_pf}
  Z^\#_{RV}&=&2\sqrt{2\pi}F_{tun}\sqrt{\frac{d^2\tilde{S}}{dE^2}}\prod_i[2\sinh(u_i/2)]^{-1}|C|^{-1/2},
  \\
  \label{ins_tun_pf}
  Z^\#_{tun}&=&\beta\sqrt{2\pi}\frac{d^2\tilde{S}}{dE^2}^{-1/2}e^{-S(\beta)}.
\end{eqnarray}
As the temperature approaches the crossover $T_c$ the effective TS
partition function $Z^\#_{RV}$ smoothly goes to the TS partition
function, Eq.~\ref{rrho_notun}, at $T>T_c$, except the
$F_{tun}\sqrt{d^2\tilde{S}/{dE}^2}$ term, cf. the discussion after
Eq.~\ref{add_tun}. Neglecting the small discontinuity related to this
term, one can consider it as a single function at all temperatures
$Z^\#_{TS}$,
\begin{eqnarray}
  \label{uni_pf}
  Z^\#_{TS}&=&Z^\#_{RV},\ T<T_c,
  \\
  \nonumber
  Z^\#_{TS}&=&Z^\#_{rot}Z^\#_{vib},\ T>T_c.
\end{eqnarray}

The expression for the effective tunneling factor $Z^\#_{tun}$,
Eq.~\ref{ins_tun_pf}, formally coincides with the expression for the
one-dimensional tunneling factor, Eq.~\ref{one_deep}, if one
understands by the one-dimensional tunneling abbreviated action the
actual, multi-dimensional one. Therefore, it can be reproduced from
Eq.~\ref{one_tun} if one assumes that the instanton effective
transimission coefficient $P(E)$ is equal to,
\begin{equation}
  \label{ins_tc}
  P(E)=e^{-\tilde{S}(-E)},\ E<0.
\end{equation}
Noticing that at higher temperatures the tunneling factor $Z^\#_{tun}$
is well reproduced by Eq.~\ref{pb_tun}, which corresponds to the
parabolic barrier transimission coefficient, Eq.~\ref{pb_tc}, and that
the abbreviated action $\tilde{S}(E)$ goes smoothly to
$2\pi E/\omega_b$ as energy approaches zero, it is then easy to see
that the synthetic transmission coefficient
$P_{TS}(E)$,
\begin{eqnarray}
  \label{uni_tc}
  P_{TS}(E) &=& \frac{1}{1 + e^{\tilde{S}(-E)}},\ E<0
  \\
  \nonumber
  P_{TS}(E) &=& \frac{1}{1 + e^{-2\pi E/\omega_b}},\ E>0.
\end{eqnarray}
can be used to obtain a tunneling factor $Z^\#_{tun}$, which goes
smoothly from low to high
temperatures. \cite{affleck81,richardson16a,kemble35} One can
calculate the integral in Eq.~\ref{one_tun} with $P_{TS}(E)$ at
temperatures above the crossover and somewhat below it (see the exact
condition below). At lower temperatures one can use Eq.~\ref{ins_rrho}
directly.

As was mentioned in the introduction a lot of effort has been devoted
to directly calculating microcanonical properties in the deep
tunneling regime using different more or less justifiable
approximations. The microcanonical properties are important both for
comparison with certain types of experimental data as well as for
subsequent theoretical applications such as the master equation
calculation for pressure dependent reactions.  It seems that the
thermal rate constant in the deep tunneling regime is much better
suited for theoretical evaluation than its microcanonical
analog. Using Eq.~\ref{uni_pf} and applying to it the inverse Laplace
transform\cite{fang21} one can calculate the ``true TS density of
states'' $\rho_{TS}(E)$, needed for the $E$-resolved rate constant
calculation. Then one can convolute it with the ``true transmission
coefficient'' $P_{TS}(E)$, Eq.~\ref{uni_tc}, to get the number of
states $N_{TS}^{(w)}(E)$, which correctly includes tunneling effects.
It is also worth noting that while each of the factors $\rho_{TS}(E)$
and $P_{TS}(E)$ has a limited physical meaning at low energies, their
combination provides the number of states, which is valid both at high
and at sufficiently low energies (but see the discussion about very
low temperatures below).

\subsection{Tunneling Effects in the Vicinity of the Crossover
  Temperature}

It is useful to estimate the tunneling effects in the vicinity of the
crossover temperature
$T_c$,\cite{affleck81,cao96,kryvohuz11,kryvohuz13,barcia14,mcconnell17}
which is, arguably, the most important temperature region for
practical applications. They mostly arise from the tunneling factor
$Z^\#_{tun}$, Eqs.~\ref{one_tun} and \ref{uni_tc}.  To this end we
expand the tunneling action up to the second order over energy $E$ in
the vicinity of the barrier top,
\begin{equation}
  \label{act_exp}
  \tilde{S}(E)=\frac{2\pi E}{\omega_b}+\alpha\frac{E^2}{\omega_b^2}+... .
\end{equation}
In further discussion we will assume that $\alpha > 0$. More
restrictions for $\alpha$ will be discussed later.

Then, the energy $E_{in}$ corresponding to the instanton trajectory can be
written as,
\begin{equation}
  \label{ins_en}
  E_{in}=\pi\alpha^{-1}\omega_b\frac{T_c-T}{T}.
\end{equation}

The main contribution to the integral in Eq.~\ref{one_tun} in the
vicinity of $T_c$ comes from the term
$\int_0^{\infty}dEe^{-\tilde{S}(E)+\beta E}$,
\begin{eqnarray}
  \label{act2}
  Z^\#_{tun}&\simeq&\beta\int_0^{\infty}e^{-\tilde{S}(E)+\beta E}dE
  \\
  \label{act2a}
            &\simeq&\beta_c\int_0^{\infty}e^{(\beta-\beta_c)E-\alpha E^2/\omega_b^2}dE,
\end{eqnarray}
where $\beta_c=2\pi/\omega_b$.  The integral in Eq.~\ref{act2a} can be
expressed via the error function
$\mbox{erf}(x)=(2\pi)^{-1/2}\int_{-\infty}^xe^{-t^2/2}dt$, giving as a
result,\cite{affleck81}
\begin{equation}
  \label{act3}
  Z^\#_{tun}\simeq2\pi^{3/2}\alpha^{-1/2}e^{\pi^2\alpha^{-1}(T-T_c)^2/T^2}
  \mbox{erf}\!\left[\sqrt{2}\pi\alpha^{-1/2}(T_c-T)/T\right],
\end{equation}
Using the asymptotic expansion of the error function,
$\mbox{erf}(x)\simeq(2\pi)^{-1/2}|x|^{-1}e^{-x^2/2},\ -x\gg1$
one obtains,
\begin{eqnarray}
  \label{act4}
  Z^\#_{tun}&\simeq&\pi^{3/2}\alpha^{-1/2},\makebox[1.4in]{}T=T_c ,
  \\
  \label{act5}
  Z^\#_{tun}&\simeq&\frac{T}{T-T_c},\makebox[1.5in]{}\frac{T-T_c}{T_c}\gg\pi^{-1}\sqrt{\alpha/2},
  \\
  \label{act6}
  Z^\#_{tun}&\simeq&2\pi^{3/2}\alpha^{-1/2}e^{\pi^2\alpha^{-1}(T-T_c)^2/T^2},\makebox[0.2in]{}
                     \frac{T_c-T}{T_c}\gg\pi^{-1}\sqrt{\alpha/2}.
\end{eqnarray}

 Equation~\ref{act5}, of course, coincides with Eq.~\ref{pb_tun}
 when
 \begin{equation}
   \label{app_con1}
 \pi^{-1}\sqrt{\alpha/2}\ll(T-T_c)/T_c\ll\pi^{-1},
\end{equation}
which is its applicability condition.  From Eq.~\ref{act6} one could
make a paradoxical conclusion that the tunneling rate constant
increases when the temperature decreases. However, one should bear in
mind that there is a much bigger decrease in the rate constant
associated with the Boltzmann factor $e^{-\beta V^\#}$ and
$Z^\#_{tun}$, with Eq.~\ref{act6} only partially compensating for
those decreases.

Actually, this provides a simple test for the applicability of the whole
approach based on the second order tunneling action expansion,
Eq.~\ref{act_exp}, in the vicinity of the crossover. Considering the
Boltzmann factor together with $Z^\#_{tun}$,
$Z^\#_{tun}e^{-\beta V^\#}$, it is reasonable to assume that this
function should decrease monotonically with the temperature.  Then
substituting into it the expression for $Z^\#_{tun}$, Eq.~\ref{act3},
and considering the so obtained expression as a function of
temperature in the vicinity of $T_c$ one obtains the following
condition on $\alpha$,
\begin{equation}
  \label{app_con3}
  \alpha^{1/2}\frac{V^\#}{\omega_b}\gg\pi^{-1/2}.
\end{equation}
This equation has a different interpretation. The parameter
$\alpha^{-1/2}\omega_b$ has a meaning of the width of the energy
window when the integral in Eq.~\ref{act2} is taken in the SPA. This
window, of course, should be much smaller than the whole integration
area, which is $V^\#$.

Equation~\ref{act6} corresponds to the SPA for the integral in
Eq.~\ref{act2} with the action given by Eq.~\ref{act_exp}. Its
applicability condition for the factor $(T_c-T)/T_c$ from below, cf.
Eq.~\ref{act6}, corresponds to temperatures for which the instanton
energy $E_{in}$, Eq.~\ref{ins_en}, is bigger than SPA energy window
$\alpha^{-1/2}\omega_b$. Meanwhile, the applicability condition for
the factor $(T_c-T)/T_c$ from above corresponds to the temperature
when the difference between the true tunneling action $\tilde{S}$ and
its expansion taken in the second order approximation,
Eq.~\ref{act_exp}, is still small in comparison with unity.  It is
difficult to estimate precisely when it is true, but if one assumes
that each successive term $\tilde{S}^{(n)}$ in the $\tilde{S}(E)$
expansion is of the order of,
\begin{eqnarray}
  \label{sp_exp}
  \tilde{S}^{(n)}\sim D^{1-n}(E/\omega_b)^n,\ D=\frac{V^\#}{\omega_b},
\end{eqnarray}
where $D^{-1}$ is a small parameter, $D\gg 1$ (see below), one would
obtain as the applicability condition for Eq.~\ref{act6},
\begin{equation}
  \label{app_con2}
  \pi^{-1}\sqrt{\alpha/2}\ll \frac{T_c-T}{T_c}\ll \alpha^{1/3},
\end{equation}
and $\alpha\sim D^{-1}$. We also note that if Eq.~\ref{sp_exp} is true,
Eq.~\ref{app_con3} is satisfied automatically.

Let us consider, as an example, the Eckart potential\cite{eckart30} with a single or
symmetric wells (both cases have the same expression for the
action). For simplicity of notations we will measure the energy and
the temperature in $\omega_b$. Then the tunneling action for the
Eckart potential is given by,
\begin{equation}
  \label{eck1}
  \tilde{S}(E)=4\pi D(1-\sqrt{1-E/D}),
\end{equation}
where $D=V^\#/\omega_b$ is the barrier height. Its expansion over $E$ is given by,
\begin{equation}
  \label{eck2}
  \tilde{S}(E)=2\pi E + \frac{\pi}{2}D^{-1}E^2+\frac{\pi}{4}D^{-2}E^3+...,
\end{equation}
where we have included also the third order term. It is worth noting
that each term in Eq.~\ref{eck2} obviously satisfies
Eq.~\ref{sp_exp}. The tunneling factor $Z^\#_{tun}$ in the vicinity of
$T_c$ can be approximated by Eq.~\ref{act2}. For clearness we will
consider $T=T_c$. Then one can see that the main contribution to the
integral
$\int_0^{\infty}dE\exp(-\frac{\pi}{2}D^{-1}E^2-\frac{\pi}{4}D^{-2}E^3-...)$
comes from the area of the width of the order of $D^{1/2}$. The third
order term in this area is of the order of $D^{-1/2}$ and is small if
$D\gg 1$. All higher order terms are sequentially smaller by the
factor of $D^{-1/2}$. Thus, the second order expansion approximation,
Eq.~\ref{act_exp}, and the SPA are justified for the Eckart barrier.
Using the SPA, Eq.~\ref{one_deep}, and the expression for
$\alpha=\frac{\pi}{2}D^{-1}$, one obtains the expression for
$Z^\#_{tun}$ for the Eckart barrier,
\begin{equation}
  \label{eck3}
  Z^\#_{tun}=2\pi^{3/2}\alpha^{-1/2}\sqrt{T/T_c}
  e^{\pi^2\alpha^{-1}(T-T_c)^2/(TT_c)},
\end{equation}
where the temperature should satisfy the inequality in Eq.~\ref{act6}.
We have written Eq.~\ref{eck3} in a form for which the correlation with
Eq.~\ref{act6} should be obvious
and the applicability condition for Eq.~\ref{act6} reads,
\begin{equation}
  \label{eck4}
  \frac{T-T_c}{T_c}\ll \pi^{-2/3}\alpha^{1/3}.
\end{equation}

Equations~\ref{act3}-\ref{act6} and \ref{eck3} may explain the
widespread use of the Eckart potential to model tunneling effects
at moderate temperatures, when the tunneling is already important but
still not too deep. One can see that, unless $\alpha$ is anomalously
small or even negative, for temperatures not too far from $T_c$, the
tunneling contribution is universal and depends only on the parameter
$\alpha$ (in addition, of course, to the tunneling frequency) for most
potentials. Then the Eckart potential becomes really handy as
it explicitly provides a simple expression for the transmission
coefficient in the energy domain and so can readily be used for
microcanonical rate constant calculations. Typically one uses the
Eckart potential that corresponds to the actual barrier height. Based
on our discussion, we suggest to use instead the Eckart potential with
the effective barrier height that corresponds to the actual value of
$\alpha$, which is obtained by some other means such as the local
potential anharmonicities.

\subsection{Connection between the Action Anharmonicity $\alpha$ and
  the Potential Expansion}

It would be useful to estimate the action anharmonicity parameter
$\alpha$, Eq.~\ref{act_exp}, in terms of the potential
anharmonicities near the barrier top,
\begin{eqnarray}
  \label{multi_pot_exp}
  V(q)&=&\frac{1}{2}\omega_b^2q_0^2-\frac{1}{2}\sum_{i=1}\omega_i^2q_i^2
  \\
  \nonumber
  &+&\frac{1}{6}\sum_{i,j,k}V^{(3)}_{i,j,k}q_iq_jq_k+\frac{1}{24}\sum_{i,j,k,l}V^{(4)}_{i,j,k,l}q_iq_jq_kq_l+....,
\end{eqnarray}
where the coordinate $q_0$ corresponds to the reaction coordinate (the
unstable mode at the saddle point) and $q_i$ correspond to stable
vibrational modes.  In Appendix E we derive an expression for
$\alpha$ in terms of the third order anharmoncities $V^{(3)}_{i,j,k}$,
\begin{equation}
  \label{alpha_m3}
  \alpha^{(3)}=\frac{\pi}{8}\left[\frac{5}{3}\frac{V_{0,0,0}^{(3)^2}}{\omega_b^5}
    -\sum_{i=1}\frac{V_{0,0,i}^{(3)^2}}{\omega_b^3}\left(2\omega_i^{-2}+\frac{1}{4\omega_b^2+\omega_i^2}\right)\right].
\end{equation}
Similarly the contribution to $\alpha$ from the fourth order
anharmonicities is given by,
\begin{equation}
  \label{alpha_m4}
  \alpha^{(4)}=-\frac{\pi}{8}\frac{V_{0,0,0,0}^{(4)}}{\omega_b^3}.
\end{equation}

Miller et al.\cite{miller90} considered the tunneling probabilties in
terms of ``good'' action variables, including effects of non-separable
coupling.  Our equations~\ref{act_exp}, \ref{alpha_m3}, and
\ref{alpha_m4} are fully consistent with their results (cf. Eqs. 13b
and 14 in Ref.\cite{miller90}, see also
Refs.\cite{greene15,greene16}). While the first term in
Eq.~\ref{alpha_m3}, which is related to the one-dimensional cubic
anharmonicity, always supresses the tunneling, the second term is
related to the curvature of the MEP, which can be considered as the
zeroth order approximation to the true tunneling
path.\cite{truhlar71,marcus77} In Appendix E we derive the expression
for the action anharmonicity correction $\alpha_{MEP}'$ in the MEP
approximation in relation to the MEP curvature in terms of the potential
cubic anharmonicity,
\begin{equation}
  \label{mep_alpha}
  \alpha_{MEP}'=-\frac{\pi}{8}\sum_{i=1}\frac{V_{0,0,i}^{(3)^2}}{\omega_b^3}
  \frac{8\omega_b^2+3\omega_i^2}{(2\omega_b^2+\omega_i^2)^2}.
\end{equation}
The difference between $\alpha_{MEP}'$ and the second term in
Eq.~\ref{alpha_m3}, which is given by,
\begin{equation}
  \label{bob_alpha}
  \alpha_{qb}'=-\frac{\pi}{2}\sum_{i=1}\frac{V_{0,0,i}^{(3)^2}}{\omega_i^2}
  \frac{\omega_b(8\omega_b^2+3\omega_i^2)}{(4\omega_b^2+\omega_i^2)(2\omega_b^2+\omega_i^2)^2},
\end{equation}
is an additional correction to $\alpha$ from the MEP
approximation. It always increases the tunneling and is related to
the so-called ``quantum bobsled'' or negative centrifugal
effect.\cite{marcus66,scodje81}.

For practical purposes it is convenient to use the anharmonic
parameters $\tilde{V}^{(3)}$ and $\tilde{V}^{(4)}$ expressed in dimensionless normal coordinates,\cite{papousek82}
in which they have the dimension of energy,
\begin{eqnarray}
  \label{m3_red}
  \tilde{V}^{(3)}_{i,j,k}&=&V^{(3)}_{i,j,k}(\omega_i\omega_j\omega_k)^{-1/2},
  \\
  \label{m4_red}
  \tilde{V}^{(4)}_{i,j,k,l}&=&V^{(4)}_{i,j,k,l}(\omega_i\omega_j\omega_k\omega_l)^{-1/2}.
\end{eqnarray}
Then Eqs.~\ref{add_tun_cub} and \ref{alpha_m3}-\ref{bob_alpha} read,
\begin{eqnarray}
  \label{add_tun_red}
  \frac{1}{2}\omega_b^2F_{tun}'&=&4\sum_i\tilde{V}^{(3)^2}_{0,0,i}\frac{\omega_b^4}{\omega_i^2}(\omega_i^2+4\omega_b^2)^{-2}\tanh(u_i/2),\
              u_i=2\pi\frac{\omega_i}{\omega_b}.
  \\
  \label{alpha_m3_red}
  \alpha^{(3)}&=&\frac{\pi}{8}\left[\frac{5}{3}\frac{\tilde{V}_{0,0,0}^{(3)^2}}{\omega_b^2}
            -\sum_{i=1}\frac{\tilde{V}_{0,0,i}^{(3)^2}}{\omega_b\omega_i}\frac{8\omega_b^2+3\omega_i^2}{4\omega_b^2+\omega_i^2}\right],
  \\
  \label{alpha_m4_red}
  \alpha^{(4)}&=&-\frac{\pi}{8}\frac{\tilde{V}_{0,0,0,0}^{(4)}}{\omega_b},
  \\
  \label{one_alpha_red}
  \alpha_{1d}&=&\frac{\pi}{8}\left[\frac{5}{3}\frac{\tilde{V}_{0,0,0}^{(3)^2}}{\omega_b^2}-\frac{\tilde{V}_{0,0,0,0}^{(4)}}{\omega_b}\right],
  \\
  \label{mep_alpha_red}
  \alpha_{MEP}'&=&-\frac{\pi}{8}\sum_{i=1}\frac{\tilde{V}_{0,0,i}^{(3)^2}}{\omega_b\omega_i}
  \frac{\omega_i^2(8\omega_b^2+3\omega_i^2)}{(2\omega_b^2+\omega_i^2)^2}.
  \\
  \label{bob_alpha_red}
  \alpha_{qb}'&=&-\frac{\pi}{2}\sum_{i=1}\frac{\tilde{V}_{0,0,i}^{(3)^2}}{\omega_b\omega_i}
  \frac{\omega_b^4(8\omega_b^2+3\omega_i^2)}{(4\omega_b^2+\omega_i^2)(2\omega_b^2+\omega_i^2)^2},
\end{eqnarray}

\begin{table}[ht]
\caption{Different approximations for the action anharmonicity
  $\alpha$ and the entanglement factor $\frac{1}{2}\omega_b^2F_{tun}'$, Eq.~\ref{add_tun_red}}
\begin{tabular}{|ccccccccc|} \hline
  Reaction
  & $\omega_b$      \textsuperscript{\emph{a}}
  & $D_{1,2}$            \textsuperscript{\emph{b}}
  & $\alpha_{Eck}$    \textsuperscript{\emph{c}}
  & $\alpha_{1d}$     \textsuperscript{\emph{d}}
  & $\alpha_{MEP}$   \textsuperscript{\emph{e}}
  & $\alpha$            \textsuperscript{\emph{f}} 
  &$\omega_b^2F_{tun}'/2$
  & Method              \textsuperscript{\emph{g}}
  \\ \hline
  $\rm CH_4+OH$
  & 1487  & 1.6, 4.5  & 0.75 & 0.44  & 0.21 & -0.67 & 0.57  & C/Q
  \\
  $\rm HNNOH(ct)$ \textsuperscript{\emph{h}}
  & 1062  & 8.1, 30.3  & 0.14 & 0.85  & 0.33 & -0.29 & 0.35  & CF/TF
  \\
  $\rm H_2CO$ \textsuperscript{\emph{i}}
  & 1841  & 16.5, 15.6  & 0.10 & 0.082  & 0.057 & 0.014 & 0.02  & C/Q
  \\
  $\rm HNNOH(tt)$ \textsuperscript{\emph{j}}
  & 1239  & 10.2, 10.4  & 0.15 & 0.4  & 0.11 & 0.074 & 0.02  & B/T
  \\
  $\rm HNNOH(tc)$ \textsuperscript{\emph{k}}
  & 1300  & 8.9, 10.7  & 0.16 & 0.36  & 0.11 & 0.081 & 0.02  & B/T
  \\
  $\rm tbuOOH$ \textsuperscript{\emph{l}}
  & 996  & 5.1, 10.7  & 0.24 & 0.13  & 0.12 & 0.1 & 0.01  & B/T
  \\
  $\rm NH_2NO$ \textsuperscript{\emph{m}}
  & 1860  & 6.2, 6.1  & 0.26 & 0.16  & 0.15 & 0.14 & $<0.01$ & B/T
  \\
  $\rm CH_3CHOO$ \textsuperscript{\emph{n}} 
  & 1603  & 3.8, 7.9  & 0.33 & 0.35  & 0.34 & 0.31 & 0.02 & B/T
  \\
  $\rm HOCO$ \textsuperscript{\emph{o}}
  & 1824  & 4.4, 5.5  & 0.32 & 0.44  & 0.43 & 0.42 & $<0.01$  & B/T
  \\
  $\rm H+H_2O_2$ \textsuperscript{\emph{p}}
  & 1289  & 1.8, 21.0 & 0.71 & 0.72  & 0.72 & 0.72 & $<0.01$  & C/Q
  \\
  $\rm C_2H_2+H$ \textsuperscript{\emph{q}} 
  & 872  & 1.6, 18  & 0.77 & 0.84  & 0.80 & 0.79 & $<0.001$  & C/Q
  \\
  $\rm NH_3+H$
  & 1594  & 2.2, 3.1 & 0.63 & 2.5  & 1.5 & 0.85 & 0.24  & C/Q
  \\
  $\rm H_2+OH$
  & 1237  & 1.7, 6  & 0.72 &  3.1 & 1.3 & 0.91 & 0.06  & C/Q
  \\
  $\rm CH_4+H$
  & 1470  & 2.9, 3.6  & 0.50 & 2.6  & 1.5 & 0.95 & 0.23  & C/Q
  \\
  $\rm H_2+H$
  & 1492  & 2.3, 2.3  & 0.69 & 4.2  & 1.9 & 1.0 & 0.29  & C/5
  \\
  $\rm CH_4+Cl$
  & 1047  & 0.6, 2.7 & 1.8 & 5.3  & 4.4 & 1.9 & 1.6  & C/Q
  \\
  $\rm H+H_2O_2$ \textsuperscript{\emph{r}}
  & 2397  & 1.4, 3.8 & 0.8 & 6.8  & 6.6 & 6.1 & 0.3  & C/Q
  \\ \hline
\end{tabular}\\
\begin{flushleft}
  \textsuperscript{\emph{a}}Imaginary frequency, 1/cm,
  \textsuperscript{\emph{b}}Barrier height $D_{1,2}=V^\#_{1,2}/\omega_b$,
  \textsuperscript{\emph{c}}$\alpha_{Eck}$, Eq.~\ref{eck_alpha}\\
  \textsuperscript{\emph{d}}$\alpha_{1d}$, Eq.~\ref{one_alpha_red},
  \textsuperscript{\emph{e}}$\alpha_{MEP}=\alpha_{1d}+\alpha_{MEP}'$, Eq.~\ref{mep_alpha_red},
  \textsuperscript{\emph{f}}$\alpha=\alpha_{MEP}+\alpha_{qb}'$,  Eq.~\ref{bob_alpha_red}\\
  \textsuperscript{\emph{g}}Methods: B=B2PLYPD3, C\{F\}=CCSD(T)\{-F12\}\\
  \makebox[0.1in]{}Basis sets: D,T,Q,5\{F\}=cc-PV\{D,T,Q,5\}Z\{-F12\}\\
  %
  %
  \textsuperscript{\emph{h}}$\rm HNNOH(ct)\rightleftharpoons N_2+H_2O$,
  \textsuperscript{\emph{i}}$\rm H_2CO\rightleftharpoons CO+H_2$,
  \textsuperscript{\emph{j}}$\rm HNNOH(tt)\rightleftharpoons HNNOH(ct)$\\
  \textsuperscript{\emph{k}}$\rm HNNOH(tc)\rightleftharpoons HNNOH(cc)$,
  \textsuperscript{\emph{l}}$\rm tbuOOH\rightleftharpoons OH+ring$,
  \textsuperscript{\emph{m}}$\rm NH_2NO\rightleftharpoons HNNOH(tc)$\\
  \textsuperscript{\emph{n}}$\rm CH_3CHOO\rightleftharpoons CH_2CHOOH$,
  \textsuperscript{\emph{o}}$\rm HOCO\rightleftharpoons CO_2+H$,
  \textsuperscript{\emph{p}}$\rm H+H_2O_2\rightleftharpoons OH+H_2O$\\
  \textsuperscript{\emph{q}}$\rm C_2H_2+H\rightleftharpoons C_2H_3$,
  \textsuperscript{\emph{r}}$\rm H+H_2O_2\rightleftharpoons H_2+HO_2$
\end{flushleft}
\label{tun_tab}
\end{table}

In Table~\ref{tun_tab} different approximations for the action
anharmoncity $\alpha$, including $\alpha_{Eck}$
for the full Eckart potential,
\begin{equation}
  \label{eck_alpha}
  \alpha_{Eck}=\frac{\pi}{2}(D_1^{-1}+D_2^{-1}-D_1^{-1/2}D_2^{-1/2}),\
  D_{1,2}=\frac{V^\#_{1,2}}{\omega_b},
\end{equation}
are shown for a set of common reactions considered in the literature.
First of all, for many reactions the Eckart potential provides a fairly
good estimate for $\alpha$. The entanglement factor
$\frac{1}{2}\omega_b^2F_{tun}'$, Eq.~\ref{add_tun_red}, is typically
small in comparison with $\alpha$ unless $\alpha$ itself is unusually
small ($\rm H_2CO$ reaction) or even negative ($\rm HNNOH(ct)$
reaction).  The one-dimensional approximation is typically bad unless the
mode coupling is small and should not be used. It is not clear so if
the higher order action expansion terms are also strongly affected by
the reaction path curvature. It is interesting to note that while the MEP
correction, Eq.~\ref{mep_alpha_red}, improves the estimate of the
action anharmonicity, it is not enough and whenever the MEP
correction is essential so also is the ``quantum bobsled'' one,
Eq.~\ref{bob_alpha_red}.

\begin{table}
\caption{Vibrational modes making more than 20\% contribution to $\alpha$}
\begin{tabular}{|ccccc|} \hline
  Reaction\textsuperscript{\emph{a}}
  & \#\textsuperscript{\emph{b}}
  & $\omega_i$\textsuperscript{\emph{c}}
  & $\alpha_{MEP}'$\textsuperscript{\emph{d}}
  & $\alpha_{qb}'$\textsuperscript{\emph{e}}
  \\ \hline
  $\rm CH_4+OH$
  &4&756&69&81\\
  $\rm HNNOH(ct)$
  &2&528&16&51\\
  &3&789&27&36\\
  &7&2269&26&2\\
  $\rm H_2CO$
  &3&1299&89&92\\
  $\rm HNNOH(tt)$
  &2&433&1&68\\
  &8&3869&98&24\\
  $\rm HNNOH(tc)$
  &2&437&1&66\\
  &8&3858&99&29\\
  $\rm tbuOOH$
  &9&429&24&33\\
  &10&493&37&38\\
  $\rm NH_2NO$
  &4&1201&7&21\\
  &7&2078&80&67\\
  $\rm CH_3CHOO$
  &2&532&6&25\\
  &4&750&16&36\\
  &14&1842&50&14\\
  $\rm HOCO$
  &2&175&75&100\\
  &5&3764&25&0\\
  $\rm H+H_2O_2$
  &2&328&4&47\\
  &4&786&11&20\\
  &5&1212&34&23\\
  &7&3797&24&1\\
  $\rm C_2H_2+H$
  &5&1924&58&74\\
  &7&3472&29&5\\
  $\rm NH_3+H$
  &6&1956&98&90\\
  $\rm H_2+OH$
  &4&2591&100&98\\
  $\rm CH_4+H$
  &8&1783&99&96\\
  $\rm H_2+H$
  &3&2053&100&100\\
  $\rm CH_4+Cl$
  &3&509&63&92\\
  &6&1194&36&8\\
  $\rm H+H_2O_2$
  &7&1533&98&94\\ \hline
\end{tabular}\\
\begin{flushleft}
\textsuperscript{\emph{a}}Reactions are listed in the same order as in
Table~\ref{tun_tab}. \textsuperscript{\emph{b}}Mode index. \textsuperscript{\emph{c}}Mode frequency, 1/cm.\\
\textsuperscript{\emph{d}}Contribution to $\alpha_{MEP}'$,
Eq.~\ref{mep_alpha_red}, in \%. \textsuperscript{\emph{e}}Contribution to $\alpha_{qb}'$,
Eq.~\ref{bob_alpha_red}, in \%.
\end{flushleft}
\label{tun1_tab}
\end{table}

Chemically, it is interesting to consider the modes that have the
largest effect on $\alpha$.  In Table~\ref{tun1_tab} we list the
vibrational modes that make major contribution to $\alpha$ for a set
of previously considered reactions (reactions are listed in the same
order as in Table~\ref{tun_tab} with the same quantum chemistry
methods). Typically, only one or very few vibrational modes make the
dominating contribution to $\alpha$. For hydrogen abstraction
reactions this mode is the mode responsible for bringing the
reactants together. It is not surprising as the corresponding
coordinate should effect the hydrogen transfer unstable frequency
$\omega_b$ the most, which may be the interpretation of the
$V^{(3)}_{0,0,i}$ term. For the same reason one can expect that values
of $\alpha$ for abstraction reactions should usually exceed those of
dissciation/recombination or internal hydrogen transfer reactions,
which is confirmed by Table~\ref{tun_tab}: The upper part of the
table, which corresponds to lower values of $\alpha$, with one
exception (the $\rm CH_4+OH$ reaction), is populated with
dissociation/recombination reactions, while its lower part includes
mostly the hydrogen abstraction reactions. The reason that the above
analysis is not applicable to the $\rm CH_4+OH\rightarrow CH_3+ H_2O$
reaction seems to be related to the extremely high anharmonical
effects in the corresponding potential, so that at least three
fundamental frequencies have negative values.

While at temperatures close to and above the crossover,
Eqs.~\ref{act3}-\ref{act6} together with Eq.~\ref{pb_tun} at higher
temperatures should provide a reasonable estimate for the tunneling
correction factor, at lower temperatures one should not expect good
agreement with experiment. The reason is that at lower temperatures
higher order terms in the tunneling action expansion start to play an
important role. Considerable effort has been devoted to calculating
the tunneling action using different corrections to the
MEP,\cite{marcus77,scodje81,scodje82,liu93,liu93a,ramos05,paneda10}
which provides a zero-order approximation to the true tunneling
path.\cite{truhlar71,marcus77} Based on our discussion, we suggest the
``local quantum bobsled'' approximation for the true tunneling
action. To this end one can calculate the tunneling action
$\tilde{S}_{mep}(E)$ along the MEP using the standard gradient
following procedure and then add to it $\alpha_{qb}'E^2/\omega_b^2$,
Eq.~\ref{bob_alpha_red}, in the hope that all higher order terms in
$\tilde{S}(E)$ expansion are included into
$\tilde{S}_{mep}(E)$. Additional research is required to test this and
other possible approximations for the ``quantum bobsled'' correction
in terms of the local properties of the potential near the barrier
top.

To summarize, Eqs.~\ref{act3}-\ref{act6} together with
Eqs.~\ref{one_alpha_red}-\ref{bob_alpha_red} provide a simple but
powerful framework to analyze, interpret, and predict experimental
results in the temperature region, where the tunneling effects are
already important but the tunneling is not too deep, which covers most
of the experimental data available. The special case of negative
$\alpha$ will be discussed next.

\subsection{Negative $\alpha$}

From Eq.~\ref{alpha_m3_red} one can see that, while for effectively
one-dimensional potentials the assumption $\alpha>0$ seems to be the
only reasonable choice, for strongly coupled multi-dimensional
potentials $\alpha < 0$ may be possible, cf., for example, the
$\rm HNNOH(ct)\rightleftharpoons N_2+H_2O$ reaction in Table~\ref{tun_tab}.
Negative values of $\alpha$ are theoretically very interesting because
in this case, at temperatures slightly above the crossover, there are
simultaneously two competing tunneling paths contributing to the rate
constant. To see that one needs to consider the next term in the
action expansion, Eq.~\ref{act_exp},
\begin{eqnarray}
  \label{act_exp_3a}
  \tilde{S}(E)&=&\frac{2\pi{E}}{\omega_b}-|\alpha|\frac{E^2}{\omega_b^2}
  +\frac{1}{3}\alpha_2\frac{E^3}{\omega_b^3}+... .
  \\
  \label{act_exp_3b}
  \frac{d\tilde{S}(E)}{dE}&=&\frac{2\pi}{\omega_b}-2|\alpha|\frac{E}{\omega_b^2}+\alpha_2\frac{E^2}{\omega_b^3}+... .  
\end{eqnarray}
When the temperature is close to but still above the crossover there
are two solutions of the $d\tilde{S}/dE=\beta$ equation,
cf. Eq.~\ref{s0},
\begin{eqnarray}
  \label{ins_en_a}
  \frac{E_{\pm}}{\omega_b}&=&\frac{|\alpha|}{\alpha_2}\pm\sqrt{\frac{\alpha^2}{\alpha_2^2}-2\pi\alpha_2^{-1}\frac{T-T_c}{T}},
  \\
  \label{ins_en_b}
\frac{1}{2}\omega_b^2\left.\frac{d^2\tilde{S}}{dE^2}\right|_{E=E_\pm}&=&\pm\sqrt{\alpha^2-2\pi\alpha_2\frac{T-T_c}{T}}.
\end{eqnarray}
The energy $E_+$ corresponds to the conventional deep tunneling
instanton, for which $d^2\tilde{S}/dE^2$ is positive. The other
solution has the shallow energy $E_-$ and the corresponding
$d^2\tilde{S}/dE^2$ is negative.  According to the conventional wisdom
the negative value of $d^2\tilde{S}/dE^2$ would correspond to the
situation when the path integral over the fluctuations arround the
instanton trajectory, Eq.~\ref{ins_pf}, has two negative
eigenvalues. Therefore, the instanton would not be a saddle point
anymore and should not contribute to the rate constant. But
if the entanglement factor $\frac{1}{2}\omega_b^2F_{tun}'$,
Eq.~\ref{add_tun_red}, exceeds the action anharmonicity $\alpha$,
which seems to be the case, this would mean that the instanton
trajectory with $E=E_-$ is still a saddle point and, therefore, does
contribute to the rate constant. It seems that the
one-dimensional approximation in terms of the thermal integral over the
transmission coefficient is not applicable in this case and the
fluctuation over energy becomes the essential part of the reaction
coordinate. We are planning to follow up on the negative $\alpha$ case
in future work.

\subsection{Example: the $\rm Cl+CH_4$ Reaction}
The reaction of Cl with CH\textsubscript{4} to produce HCl and
CH\textsubscript{3} provides a simple case upon which to test the
suitability of the methods derived here via comparison with experiment.
The experimental data for this reaction extends down as low as 181 K in
the forward direction\cite{seely96} and 187 K in the reverse direction,\cite{eskola06} which is
significantly lower than the estimated critical temperature of 230 K.
Furthermore, there is good consistency in the measurements across
multiple labs over a many year time period. Just as importantly, the
small size of the reactants allows for the use of high levels of
electronic structure theory in investigating the key molecular
parameters. As a result, the uncertainties in other aspects of the
analysis are small enough that a comparison with experiment is primarily
indicative of our ability to predict the effects of tunneling.

The explicitly correlated CCSD(T)-F12 method\cite{adler07,knizia09} was used with the
cc-pVQZ-F12 basis set\cite{peterson08} to predict the stationary point geometries and
vibrational frequencies. Higher accuracy estimates for the stationary
point energies were obtained from a composite approach incorporating (i)
CCSD(T) estimates for the CBS limit based on explicit calculations for
the aug-cc-pVnZ series with n=5 and 6,\cite{dunning89} (ii) CCSDT(Q)/cc-pVTZ and
CCSDTQ(P)cc-pVDZ corrections for the effect of higher order excitations,
(iii) CCSD(T)/CBS calculations of core-valence effects based on
extrapolations of results for the cc-pCVTZ and cc-pCVQZ basis sets, (iv)
CCSD(T)/aug-cc-pwCV5Z-DK calculations of scalar relativistic effects
employing the Douglas-Kroll-Hess (DKROLL=1) one electron integrals,\cite{peng12} (v)
CCSD/cc-pVTZ evaluations of the diagonal Born-Oppenheimer correction
(DBOC), (vi) CCSD(T)-F12/cc-pVQZ-F12 zero point energies, and (vii)
B2PLYP-D3/cc-pVTZ\cite{grimme10} based calculations of anharmonic vibrational
corrections through second order vibrational perturbation theory. For
the Cl + CH\textsubscript{4} reactants, the experimentally determined
spin-orbit lowering was added to the energy. For all other stationary
points, the spin-orbit splitting was presumed to be zero.

The calculations were performed with MOLPRO,\cite{molpro15,werner12}
except for the evaluations of the DBOC, which was performed with
CFOUR.\cite{cfour} The CCSDT(Q) calculations employed Kallay's MRCC
extension to MOLPRO.\cite{mrcc,kallay01} The coupled cluster
calculations employed restricted spin Hartree-Fock wavefunctions
within the unrestricted coupled cluster formalism, except for the
CCSDT(Q) calculations which employed unrestricted spin
wavefunctions for both the HF and coupled-cluster components. The
CCSDT(Q) correction was then taken as the difference between the
UUCCSDT(Q)/cc-pVDZ result and the RUCCSD(T)/cc-pVDZ result.

 The results of these calculations of the energies are reported in
Table~\ref{cl_ch4_tab}. Notably, the predicted reaction energy of
1.19 kcal/mol for this composite approach agrees with the Active
Thermochemical Tables value of 1.15 +/- 0.02 kcal/mol,\cite{atct}
which is within the expected uncertainty. The zero-point corrected
barrier height in the forward and reverse directions were predicted to
be 3.67 and 2.48 kcal/mol, respectively.

\begin{table}
\caption{Stationary Point Energies (kcal/mol) on the 
$\rm CH_4Cl$ Potential Energy Surface\textsuperscript{\emph{a}}}
\begin{tabular}{|ccccccccccccc|} \hline
  Species
  &\multicolumn{4}{c}{CCSD(T)}&T(Q)& Core&Rel
  &DBOC&Har&Anh&SO&Total\textsuperscript{\emph{b}}
  \\
  &AQZ&A5Z&A6Z&CBS&TZ&Val&&&\multicolumn{2}{c}{ZPE}&&
  \\ \hline
  $\rm CH_4+Cl$
  &0&0&0&0&0&0&0&0&0&0&0&0
  \\
  $\rm CH_4\ldots Cl$
  &-0.38&-0.36&-0.34&-0.31&-0.01&0.00&0.01&0.02&1.20&&0.84\textsuperscript{\emph{c}}&1.74
  \\
  $\rm Cl\ldots CH_4$
  &-0.99&-0.99&-0.97&-0.95&-0.01&0.00&0.01&0.11&0.21&&0.84\textsuperscript{\emph{c}}&0.20
  \\
  TS
  &7.16&6.92&6.94&6.97&-0.16&0.05&0.21&0.08&-4.25&-0.07&0.84\textsuperscript{\emph{c}}&3.67
  \\
  $\rm CH_3\ldots HCl$
  &3.22&2.81&2.80&2.78&-0.05&-0.02&0.34&0.00&-3.87&0.39&0.84&0.41
  \\
  $\rm CH_3+HCl$
  &5.35&5.04&5.04&5.04&-0.03&-0.03&0.31&0.01&-5.14&0.19&0.84&1.19
  \\ \hline
\end{tabular}\\
\begin{flushleft}
\textsuperscript{\emph{a}}All at the CCSD(T)-F12/cc-pVQZ-F12
optimized geometry.\\
\textsuperscript{\emph{b}}Q(P)/DZ correction is less than 0.01
kcal/mol for all configurations.\\
\textsuperscript{\emph{c}}These values are based on a presumption that there is no spin-orbit coupling for anything other than the reactants.
\end{flushleft}
\label{cl_ch4_tab}
\end{table}

The cubic and quartic force constants were evaluated at the
CCSD(T)/cc-pVQZ level. The calculations yield a predicted second order
action term, $\alpha$, of 1.91, while the EF is predicted to be
1.55. Notably, this EF is the largest we have calculated. It
corresponds to a reduction in the predicted rate constant by a factor
of 1.36. The MEP and quantum bobsled effects have a significant effect
on the overall $\alpha$ value. The one-dimensional $\alpha_{1d}$ is
much larger (5.3), with contributions of -0.9 and -2.5 from the MEP
and quantum bobsled effects.

\begin{figure}[p]
\begin{center}
\includegraphics*[height=8in, width=6in]{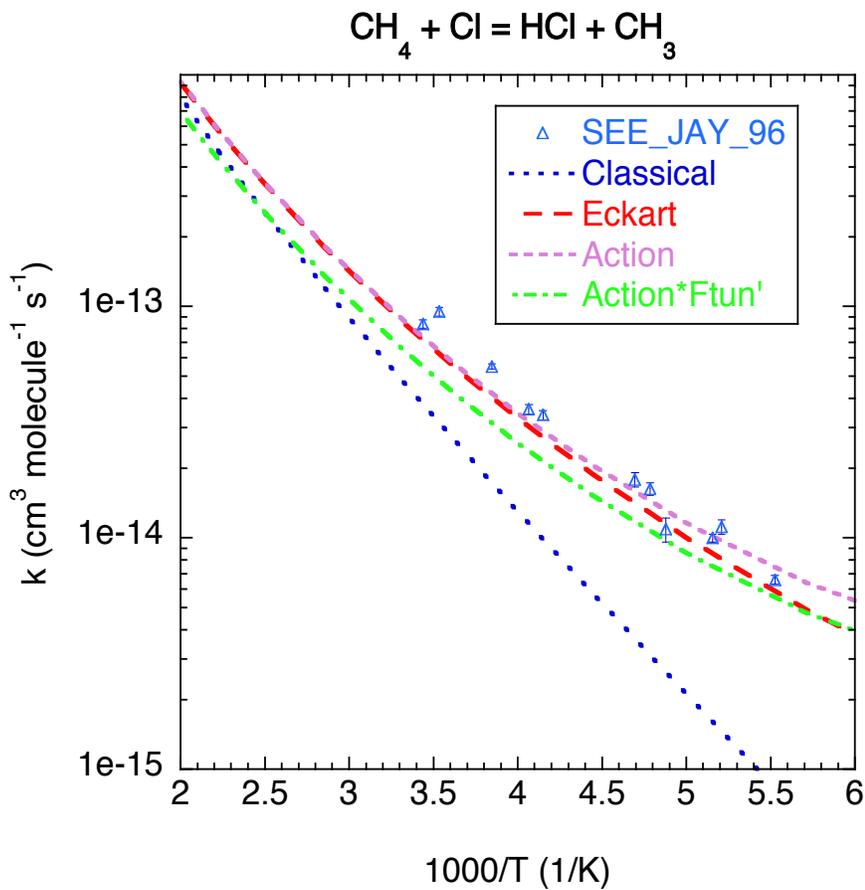}
\caption{
Plot of the temperature dependence of the thermal rate
constants the $\rm Cl + CH_4\rightarrow HCl + CH_3$
reaction. The present results based on the evaluation of the tunneling
probability for the quadratic expansion of the action are denoted with
green dashed-dotted and purple dashed lines, for calculations with and
without the EF included, respectively.
}
\label{ch4_cl_fig}
\end{center}
\end{figure}

\begin{figure}[p]
\begin{center}
\includegraphics*[height=8in, width=6in]{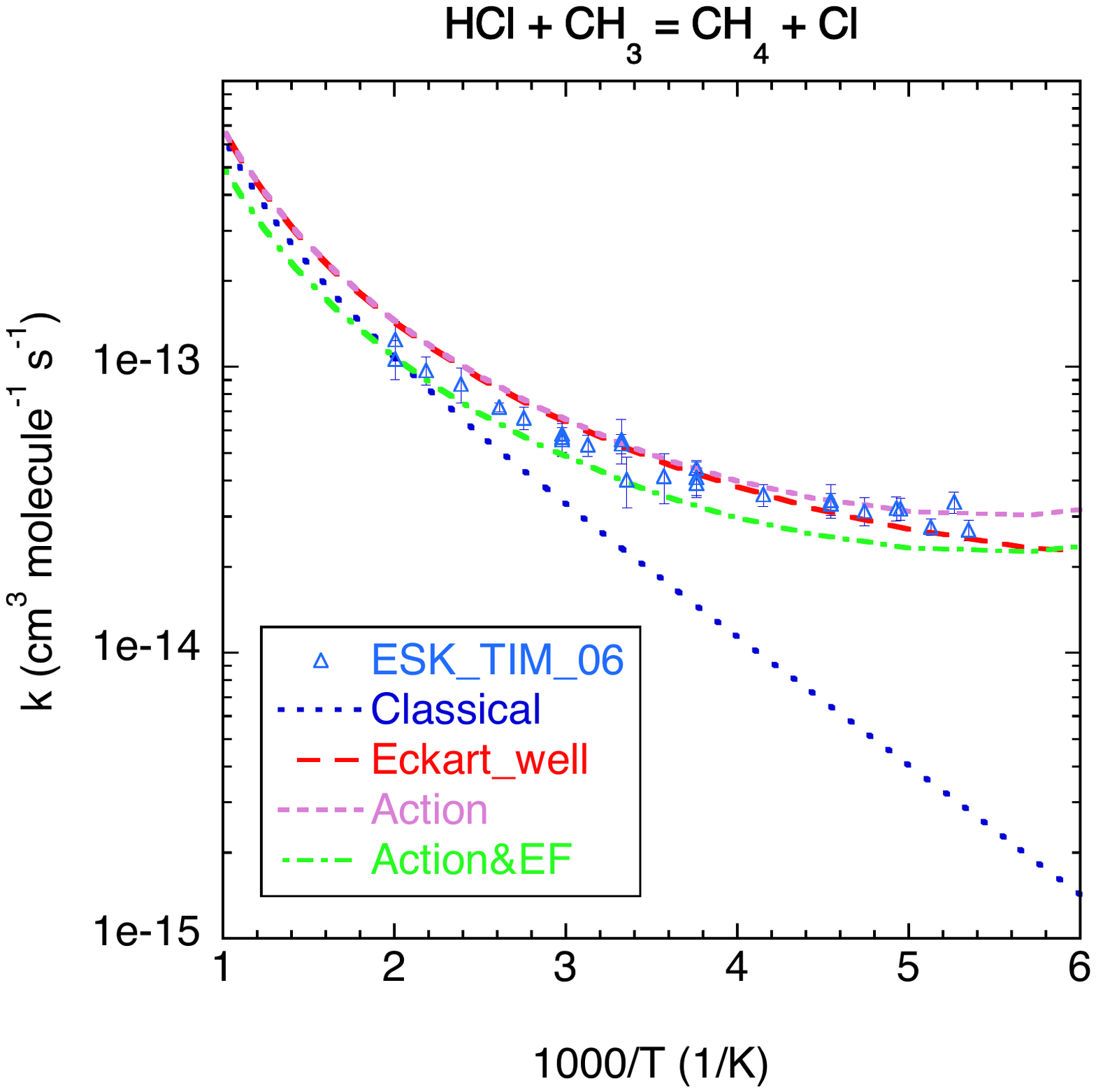}
\caption{
Plot of the temperature dependence of the thermal rate
constants for the $\rm HCl + CH_3\rightarrow Cl + CH_4$
reaction. The present results based on the evaluation of the tunneling
probability for the quadratic expansion of the action are denoted with
green dashed-dotted and purple dashed lines, for calculations with and
without the EF included, respectively.
}
\label{ch3_hcl_fig}
\end{center}
\end{figure}

 The rate constants were calculated according to Eq.~\ref{rrho}
employing CCSD(T)-F12/cc-pVQZ-F12 calculated rovibrational properties
to evaluate $Z_{rot}^\#$ and $Z_{vib}^\#$ within the RRHO
approximation. The tunneling factor $Z_{tun}^\#$ was evaluated via
direct integration of Eq.~\ref{one_tun} employing Eq.~\ref{uni_tc} to
represent the tunneling probability.  The action is evaluated with the
quadratic expansion in energy. The integral in Eq.~\ref{one_tun} is
truncated at the bottom of the well.

The resulting predictions for the forward and reverse thermal rate
constants are compared to the two lowest temperature experimental
studies for the forward and reverse reactions in
Figs.~\ref{ch4_cl_fig} and \ref{ch3_hcl_fig}. For comparison purposes,
we illustrate results with and without the EF, classical results
(i.e., with $Z_{tun}^\#$ set to unity), and Eckart tunneling
predictions. These Eckart predictions employ the zero-point corrected
higher order energy estimates of the barrier relative to the van der
Waals prereactive complexes, which yields a slightly lower $\alpha$
value of 1.4. Note though that the analytic expression for the Eckart
tunneling factor is used here in place of the action expansion.

At the lowest temperatures tunneling is seen to increase the rate
coefficient by about an order of magnitude. The present quadratic action
based perturbative multi-dimensional tunneling approximation accurately
captures this effect, with the experimental results generally falling
between the results with and without the EF correction. The results for
the reverse direction are in slightly better agreement with the
experimental data, suggesting that there may be some minor inadequacy in
the exothermicity/barrier calculation. Overall, the level of agreement
is quite remarkable for a fully a priori calculation of the rate
constants. The Eckart predictions are also in good agreement with the
experimental data, which, as has been observed previously, is at least
in part the result of some fortunate cancellation of errors.

\subsection{Higher Order Expansion for One-Dimensional Problems}

In some situations multi-dimensional effects are not important and
the one-dimensional approximation provides a reasonable
estimate of the value of $\alpha$ for a particular problem. We
derive the corresponding expressions using a different method in
Appendix F,
\begin{equation}
  \label{alpha_res}
  \alpha =\frac{\pi}{8}\left(\frac{5}{3}V_3^2-V_4\right),
\end{equation}
where we have used a slightly different representation for the
potential expansion, Eq.~\ref{one_pot_exp}.  As it should be, $V_3$
always has a positive contribution to $\alpha$ and $V_4$ contributes
with the opposite sign.  From our derivation it is clear that $V_3$
and $V_4$ are the only terms in the one-dimensional potential
expansion that contribute to $\alpha$. For reference purposes, we also
provide the next tunneling action expansion coefficient $\alpha_2$,
cf.  Eq.~\ref{act_exp_3a}, in terms of the potential anharmonicities,
Eq.~\ref{one_pot_exp}, including the $120^{-1}V_5\omega_b^{7/2}q^5$
and $720^{-1}V_6\omega_b^4q^6$ terms,
\begin{equation}
  \label{alpha2}
 \frac{\alpha_2}{\pi}=-\frac{1}{48}V_6+\frac{7}{48}V_3V_5
+\frac{35}{384}V_4^2-\frac{35}{64}V_3^2V_4+\frac{385}{1152}V_3^4,
\end{equation}
which has been obtained from Eqs.~\ref{act_com1} and \ref{act_com2}. 

It is interesting to find a potential, for which the action expansion
in Eq.~\ref{act_exp} is actually limited by the first two terms. In
Appendix F it is shown that such a potential is given by the implicit
equation,
\begin{equation}
  \label{act_mod6}
  q=y+\alpha\frac{\omega_b}{3\pi}y^3,\ V(q)=\frac{1}{2}\omega_b^2y^2.
\end{equation}
One can see that at small $y$, which correspond to small $q$, the
potential is quadratic, $V(q)\propto q^2$, as it should be. At larger
$y$, which correspond to larger $q$, the potential is flattened out and
becomes proportional to $q^{2/3}$, $V(q)\propto q^{2/3}$.

\subsection{Low Temperature Limitations}

\begin{figure}[p]
  \begin{center}
    \includegraphics*[height=8in, width=6in]{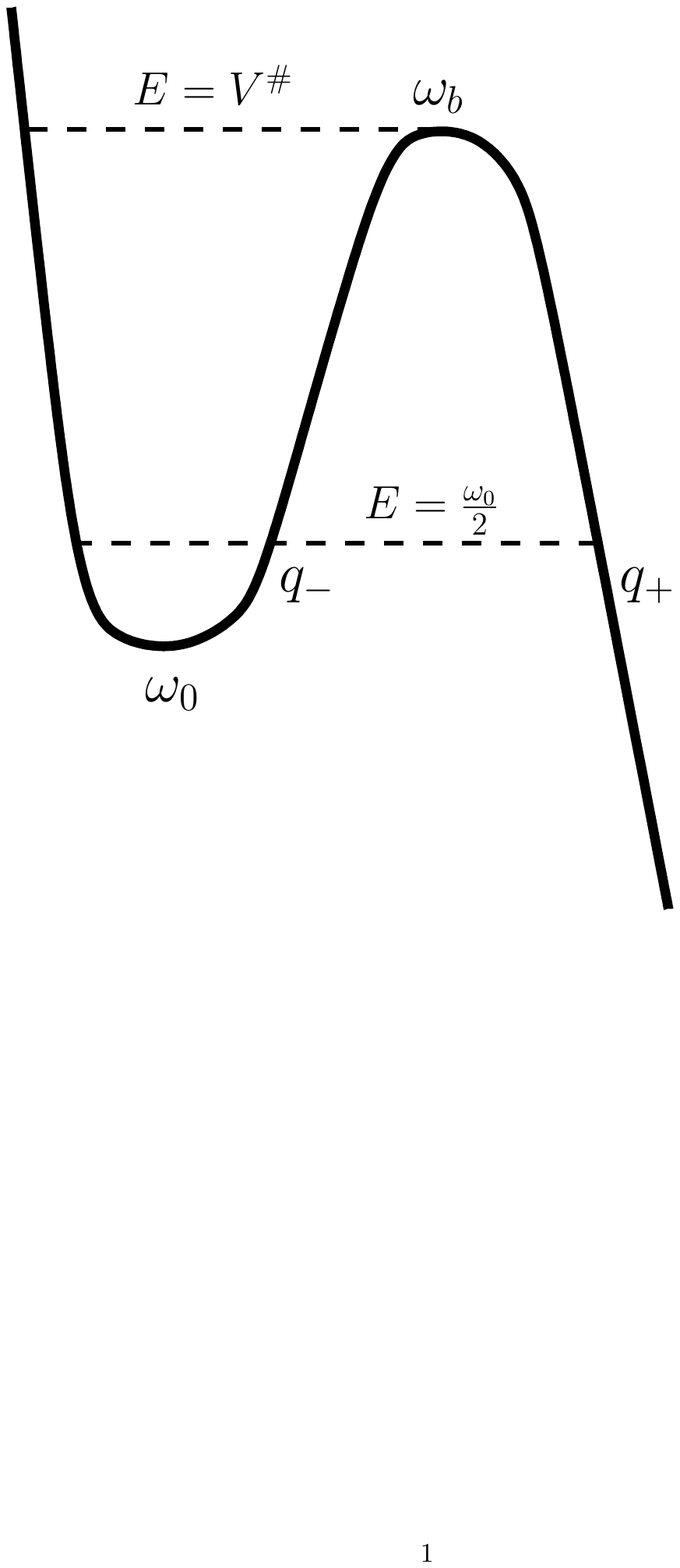}
\caption{one-dimensional metastable potential}
\label{fig_pot}
\end{center}
\end{figure}

While for bimolecular reactions Eq.~\ref{one_tun} is valid at all
temperatures, for unimolecular decomposition at low temperatures it is
not true. This can be seen already for the one-dimensional metastable
potential, Fig.~\ref{fig_pot}.
In Appendix G, it is shown that at very low temperatures, when the
instanton trajectory is close to the bottom of the well, the
abbreviated tunneling action is approximately equal to,
\begin{equation}
  \label{lt_act}
\tilde{S}(V^\#-E)=\tilde{S}^\#+\frac{E}{\omega_0}\ln\left(\frac{E}{eE_0}\right),\ 
  \tilde{S}^\#=\tilde{S}(V^\#) ,
\end{equation}
where $\omega_0$ is the harmonic frequency at the bottom of the well,
$e$ is the natural logarithm base, and $E_0$ is the characteristics of
the tunneling barrier, cf. Eq.~\ref{f4}. In contrast with our previous
consideration, the energy $E$ in Eq.~\ref{lt_act} is counted from the
bottom of the well upward. Substituting Eq.~\ref{lt_act} into
Eq.~\ref{uni_tc} or \ref{wkb_tc} and then integrating over $E$
according to Eq.~\ref{one_tun}, one obtains the expression for for
$Z^\#_{tun}\simeq \exp(-\tilde{S}^\#+\beta V^\#)$ at very low
temperatures, $\beta\omega_0\gg1$, which has an obviously wrong
temperature dependence: The reactant partition function, to which only
the ground state contributes, and the Boltzmann factor give together
$\exp[-\beta (V^\#-\omega_0/2)]$ and so the rate constant at very low
temperatures would read,
$k\simeq(2\pi\beta)^{-1}\exp(-\tilde{S}^\#+\beta\omega_0/2)$, while
the true rate constant, which corresponds mostly to the tunneling from
the ground metastable state, should approach a constant. In contrast,
as shown in Appendix G, Langer's theory, for which the expression in
the one-dimensional case formally corresponds to the SPA,
Eq.~\ref{one_deep}, for the integral in Eq.~\ref{one_tun}, gives the
correct result.

\begin{figure}[p]
\begin{center}
\includegraphics*[height=8in, width=6in]{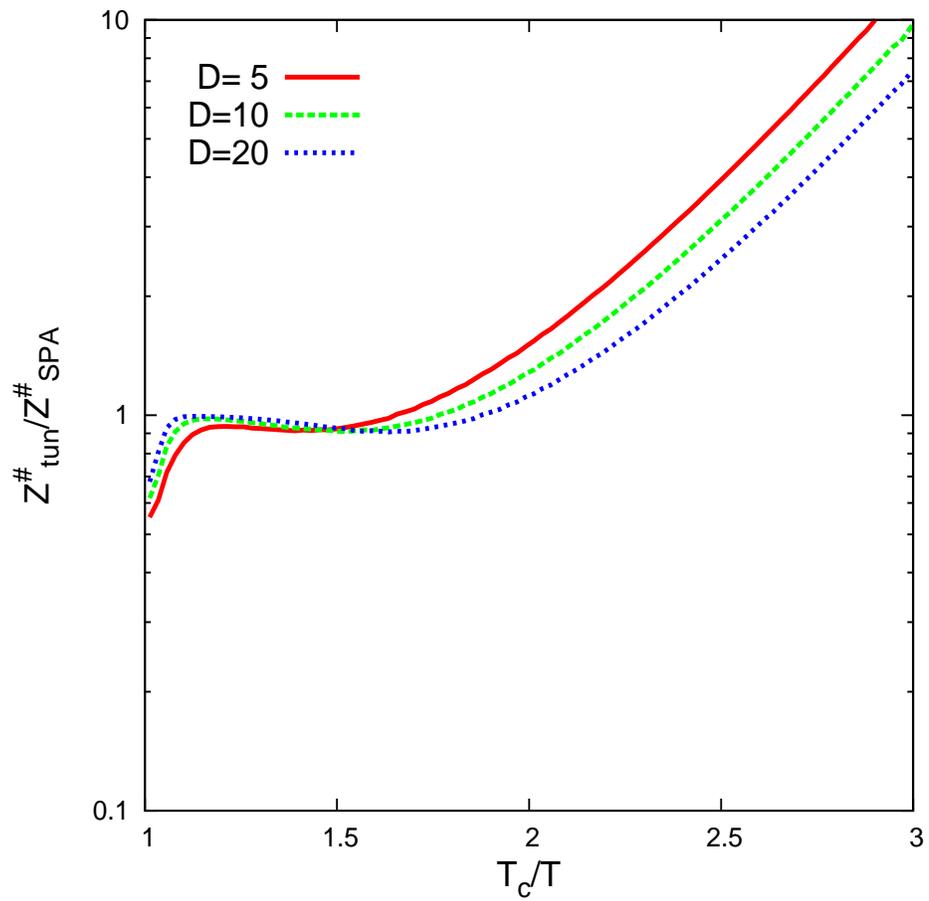}
\caption{The ratio of the tunneling factor $Z^\#_{tun}$,
  Eq.~\ref{one_tun}, to its SPA approximation as a function of
  temperature for different values of the barrier height $D=V^\#/\omega_0$ for the
  cubic potential, Eq.~\ref{cub_pot}}
\label{fig_cub}
\end{center}
\end{figure}

The mathematical reason for this behavior is that the SPA energy
window in the integral in Eq.~\ref{one_tun}, cf. Eq.~\ref{f13} in
Appendix G, goes far beyond the upper limit of integration and,
therefore, the SPA is formally unapplicable.  In Fig.~\ref{fig_cub} we show
the ratio of the tunneling factor $Z^\#_{tun}$ obtained on the basis
of Eq.~\ref{one_tun} and in the SPA, Eq.~\ref{one_deep}, as a function of
temperature for the cubic potential,
\begin{equation}
  \label{cub_pot}
  V(q)=\frac{1}{2}\omega_0^2q^2-\frac{1}{6}\sqrt{\frac{2}{3}}\omega_0^{5/2}D^{-1/2}q^3,
\end{equation}
where $D=V^\#/\omega_0$ is the barrier height in frequency units (for
a cubic barrier $\omega_b=\omega_0$).
This ratio basically shows the error that one makes when one directly
evaluates the integral over energy in Eq.~\ref{one_tun}
(for the one-dimensional or the multi-dimensional potential, does not
matter) instead of using the SPA.

One may think that the expression with the integral over energy for
the tunneling factor, Eq.~\ref{one_tun}, is more accurate than the SPA and
that the SPA is just an approximation to it.  The physical reason of
inapplicability of Eq.~\ref{one_tun} at very low temperatures is that
it is based on the assumption that the dynamics in the well and the
tunneling dynamics are independent processes. In the foundation of
Eq.~\ref{one_tun} lies the assumption that one can describe the
tunneling effect in terms of the transmission coefficient for the
incoming and outgoing semiclassical waves with an arbitrary
energy. While at higer temperatures and, correspondingly, energies
this is a reasonable assumption, for low temperatures this is not true
at all. Only well-defined metastable states with certain energies do
contribute to the rate constant. Therefore, to provide the expression
analogous to Eq.~\ref{one_tun} at low temperatures, one has to
consider the tunneling from individual quantum states, which is a very
difficult task in the multi-dimensional, non-separable case.

In contrast Langer's theory does not consider tunneling with certain
energies but rather considers the thermal process from the
beginning. It is important to stress that the tunneling trajectory in
Langer's theory does not correspond to any real quantum state energy
and is simply a mathematical construct.  At relatively high energies
when the density of states is sufficienty high, it corresponds to the
energy region that gives the main contribution to the tunneling rate
constant. At very low temperatures the energy of the tunneling
trajectory is exponentially close to the bottom of the potential, see
Eq.~\ref{f12} in Appendix G, and has nothing to do with the ground
state energy.

\subsection{Numerical Implementation}

For practical numerical applications of Eqs.~\ref{multi_ins}, \ref{tun_pref},
and \ref{ins_rrho} one must find the instanton trajectory as well as
the stability parameters and the matrix $C_{\xi,\eta}$ for each
considered energy.\cite{liberto16,mcconnell17a,winter19} The instanton
trajectory search is facilitated by the following considerations.  One
can start from the vicinity of the saddle point where the instanton
trajectory is well known and corresponds to the motion along the
unstable vibrational mode, the vectors $q_i$ and $p_i$ correspond to
the transverse normal modes vectors with the appropriate
normalization, and the stability parameters are
$u_i=2\pi\omega_i/\omega_b$.  For each subsequent energy, one can use
the linearized equations of motion, Eq.~\ref{ins_lem}, to obtain the
monodromy transformation matrix $\hat{M}$ and then to obtain from it
the new vectors $q_i$, $p_i$, and stability parameters $u_i$.  To
obtain $q_E$, which is the derivative of the instanton trajectory
turning point with respect to the energy, one can use Eqs.~\ref{ort_rel2},
\ref{ort_rel3}, and \ref{ort_rel9} and express the vector $q_E$ as a
linear combination of the vectors $p_i$ and $g_0$.  The ``inertia
moments matrix'' $C_{\xi,\eta}$, Eq.~\ref{jv_mon1}, can be obtained by
applying the monodromy transformation to the vectors $v^J_\xi$,
Eq.~\ref{am_dyn}, and considering the $q$-part of the resulting
vectors.  To make the next step by going down in energy, a new turning
point can be used as a starting guess for the new trajectory. The new
turning point is obtained by going along the vector $q_E$ and can be
more accurately adjusted by shooting the trajectory and using the
previous vectors $q_i$ and $p_i$ and the stability parameters for
adjustment.  It is interesting to note that in general $q_E$ does not
coincide with (but is close to) $g_0$, which is the energy gradient in
mass-weighted coordinates. Otherwise, the instanton trajectory turning
point as a function of energy would coincide with the minimum energy
path. This deviation is the manifestation of the coupling between the
reaction coordinate and the transverse modes.

\subsection{Anharmonicity in Low Frequency Modes}

Sufficiently large chemical complexes usually have many modes
including low-frequency ones. For such systems the second-order
harmonic approximation, which is inherently present in Langer's
theory, may be too restrictive because for these modes considerable
anharmonical corrections may be required even below the crossover
temperature. To handle this situation one can use an {\em ad hoc}
approach in which the full TS partition function is written in the
following form,\cite{beyer16}
\begin{equation}
  \label{an_pf}
  Z^\#\simeq\frac{Z^\#_{AH}}{Z^\#_{RRHO}}Z^\#_{ins},
\end{equation}
in which $Z^\#_{AH}$ is the anharmonic partition function for the TS
with no tunneling, $Z^\#_{RRHO}$ is the partition function above the
barrier in the harmonic approximation, Eq.~\ref{rrho_notun}, and
$Z^\#_{ins}$ is the partition function with tunneling from
Eq.~\ref{ins_rrho}.  Equation~\ref{an_pf} provides the correct
expression for the partition function at high and low temperatures. If
some anharmonic low frequency modes are not involved in the tunneling
dynamics (i.e. they are uncoupled from the reaction coordinate) they
naturally come into Eq.~\ref{an_pf} as a factor and the
corresponding harmonic terms are automatically cancelled out.

\subsection{Correspondence between Negative Eigenvalues and Focal
  Points of the Instanton Trajectory}

The instanton trajectory corresponds to a saddle point in the
configurational space of closed paths. This feature means that there
should be only one negative eigenvalue in the quadratic action
$S'[\delta q]$, Eq.~\ref{ins_act}, considered for small deviations
from the instanton trajectory. On the other hand, the path integral
considered as a function of the final time is inversely proportional
to the square root of the certain determinant $|q_1q_2...q_N|$,
cf. Eq.~\ref{pref}. When the tunneling trajectory passes through the
focal point this determinant becomes zero and the path integral
diverges, which means that one more eigenvalue in the quadratic action
$S'[\delta q]$ has passed through zero. The determinant after passing
the focal point changes sign. Thus, the number of focal points of the
tunneling trajectory corresponds to the number of negative eigenvalues
in the quadratic action $S'[\delta q]$ and should be equal 1.

In the derivation of the rate constant expressions we assumed that the
turning points for the instanton trajectory, in which the system comes
to complete rest, $\dot{q}_0=0$, exist. We would like to give an
argument in support of this assumption.  From classical mechanics
it is known that the minimization of the classical action,
cf. Eq.~\ref{act_rel}, can be equivalently reformulated in the form of
Maupertuis's principle.\cite{arnold78} Let us consider the curve that
connects two hypersurfaces of constant energy on both sides of the
barrier separating reactants and products and that it is parametrized so as
to satisfy $E= \dot{q}^2/2 + V(q)$ condition. Then the true classical
trajectory corresponds to the minimum of the abbreviated action,
$\tilde{S}[q]$,
\begin{equation}
  \label{maup}
  \tilde{S}=\int \dot{q}dq = \int \dot{q}^2 d\tau,\ E=\dot{q}^2/2+V(q).
\end{equation}
It is obvious that such a trajectory always exists and the points
where it touches the $E=V(q)$ hypersurfaces are the turning
points. Returning to the regular description and considering the
variations of the turning points for the half-trajectory one concludes
that this trajectory corresponds to the extremum of the tunneling
action, Eq.~\ref{tun_act}.

\section{Concluding Remarks}

The tunneling effects for general, non-separable systems have been
considered within the framework of Langer's theory. A closed form rate
expression has been obtained, which differs from the commonly used one
by an additional term related to the system non-separability. We have
also obtained a general rate expression that takes into account
angular momentum conservation. A simple model has been suggested to
describe the tunneling effects in the vicinity of the crossover
temperature, which is based on the local potential properties near the
saddle point.

\newpage
\appendix{\bf\Large Appendices}

\renewcommand{\thesection}{\Alph{section}.}
\renewcommand{\theequation}{\Alph{section}.\arabic{equation}}

\setcounter{equation}{0}
\section{Derivation of the Prefactor $U(t)$ for the Harmonic Path Integral}
In this section we use the notations $S'(q, t)=S'(0,q,t)$, and
$K(q, t)=K(q,0,t)$.  Then for small time steps $\delta t$ the time
dependence of the prefactor $U(t)$ for the harmonic path integral can
be obtained from,
\begin{equation}
  \label{a1}
  U(t + \delta t) = U(t)(2\pi\delta t)^{-N/2}\int d^Nq
  e^{-S'(q,t)}e^{-q^2/2\delta t},
\end{equation}
where we have used the expression for $K(q, \delta t)$ at small
$\delta t$,
\begin{equation}
  \label{a2}
  K(q,\delta t)=(2\pi\delta t)^{-N/2}e^{-q^2/2\delta t}.
\end{equation}
The action $S'(q,t)$ can be written as
$S'(q,t)=q(t)\dot{q}(t)/2$. Because $q(t)$ satisfies linear
equations of motion, $\dot{q}(t)$ is a linear function of $q$,
\begin{equation}
  \label{a3}
  \dot{q}(t)=\hat{G}(t)q.
\end{equation}
Substituting Eq.~\ref{a3} into Eq.~\ref{a1} and expanding the exponent
$e^{-S'(q,t)}$ up to first order in $q$ (higher order terms
correspond to
higher orders in $\delta t$), one obtains,
\begin{equation}
  \label{a4}
  U(t + \delta t) = U(t)(1 -\frac{1}{2}\delta t\sum_iG_{i,i}),
\end{equation}
On the other hand, considering the determinant 
$D(t)=|q^{(1)}(t),q^{(2)}(t),...,q^{(N)}(t)|$  for $N$ trajectories, which start
with some initial velocities and zero coordinates, one finds, that
\begin{eqnarray}
  \label{a5}
  \frac{dD}{dt}&&=|\dot{q}^{(1)},q^{(2)},...,q^{(N)}|+|q^{(1)},\dot{q}^{(2)},...,q^{(N)}|
  \\
  \nonumber
  &&+...+|q^{(1)},q^{(2)},...,\dot{q}^{(N)}|.
\end{eqnarray}
Considering the velocities $\dot{q}^{(i)}(t)$ in the basis of $q^{(i)}(t)$,
$\dot{q}^{(i)}(t)=\tilde{G}_{j,i}(t)q^{(j)}(t)$, one finds that
$\frac{dD}{dt}=D\sum_i\tilde{G}_{i,i}$.
But $\hat{\tilde{G}}=\hat{Q}^{-1}\hat{G}\hat{Q}$, where
$Q_{i,j}=q_i^{(j)}(t)$, and the trace does not depend on the
basis. Therefore,
\begin{equation}
  \label{a6}
  \frac{dD}{dt} = \sum_iG_{i,i}D.
\end{equation}
Substituting Eq.~\ref{a6} into Eq.~\ref{a4} one obtains,
\begin{equation}
  \label{a7}
  \frac{1}{U}\frac{dU}{dt}= -\frac{1}{2}\frac{1}{D}\frac{dD}{dt},
\end{equation}
whose solution is $U(t)=CD^{-1/2}$. Choosing the constant $C$ to match
the determinant $D$ at small times $t$,
$D(t)\simeq|t\dot{q}^{(0)}(0),t\dot{q}^{(1)}(0),...,t\dot{q}^{(N)}(0)|$, and taking into
account that $U(t)$ at small times is given by,
$U(t)\simeq(2\pi t)^{-N/2}$, cf. Eq.~\ref{a2}, one arrives at
Eq.~\ref{pref}.

\setcounter{equation}{0}
\section{Perturbation Theory for $U(T)$}

We will calculate the determinant ratio R, cf. Eq.~\ref{pref}
\begin{equation}
  \label{ratio}
  R = \frac{|q^{(1)}(T+\tau),q^{(2)}(T+\tau),...,q^{(N)}(T+\tau)|}{|p^{(1)}(\tau),p^{(2)}(\tau),...,p^{(N)}(\tau)|},
\end{equation}
in the leading order of $\tau$, which happens to be two, and we will use the
notation $p=\dot{q}$. First one notes that the monodromy matrix is left
unchanged with the time-shift $\tau$ if one uses for the shifted
matrix the basis,
\begin{equation}
  \label{b1}
\tilde{v}_i=\hat{K}(\tau,0)v_i,
\end{equation}
where $\hat{K}(t_2,t_1)$ is the time-propagator for the linearized
equations of motion, Eq.~\ref{ins_lem}. From the definition of the
monodromy matrix it follows that
\begin{equation}
  \label{b2}
  \hat{K}(T,0)v_i=\hat{M}_{i,j}v_j.
\end{equation}
Multiplying Eq.~\ref{b2} by $\hat{K}(\tau,0)$ one arrives at
\begin{equation}
  \label{b3}
  \hat{K}(T+\tau,\tau)\tilde{v}_i=\hat{M}_{i,j}\tilde{v}_j,
\end{equation}
where we have used the periodicity of the tunneling trajectory, $\hat{K}(T+\tau,T)=\hat{K}(\tau,0)$.

Let us calculate $\hat{K}(\tau, 0)$ up to the second order in
$\tau$. From Eq.~\ref{ins_lem} it follows, that $\ddot{p} =-\hat{k}p$,
where $$\hat{k}=\frac{\partial^2V}{\partial q^2}$$ and we have used the
fact that $\dot{\hat{k}}=\frac{\partial^3V}{\partial q^3}\dot{q}_0=0$ at
the turning point. Therefore, the time-shifted coordinates and momenta
read up to second order, 
\begin{eqnarray}
  \label{b4}
  q(\tau)&\simeq&q+\tau p-\frac{1}{2}\tau^2\hat{k}q,\\
  p(\tau)&\simeq&p-\tau\hat{k}q-\frac{1}{2}\tau^2\hat{k}p, \nonumber
\end{eqnarray}
or in matrix form,
\begin{equation}
  \label{b5}
  \hat{K}(\tau,0)\simeq\hat{I}+
  \tau\left(\begin{array}{cc}0&I\\-\hat{k}&0\end{array}\right)+
  \frac{1}{2}\tau^2\left(\begin{array}{cc}-\hat{k}&0\\0&-\hat{k}\end{array}\right).
\end{equation}
We will also need $\hat{K}(\tau,0)^{-1}$, which is given by,
\begin{equation}
  \label{b6}
  \hat{K}(\tau,0)^{-1}\simeq\hat{I}-
  \tau\left(\begin{array}{cc}0&I\\-\hat{k}&0\end{array}\right)+
  \frac{1}{2}\tau^2\left(\begin{array}{cc}-\hat{k}&0\\0&-\hat{k}\end{array}\right),
\end{equation}
where we have used the identity
$(\hat{I}+\tau\hat{A}+\tau^2\hat{B})^{-1}\simeq\hat{I}-\tau\hat{A}-\tau^2\hat{B}+\tau^2\hat{A}^2$.

We will use the the monodromy matrix eigenspace representation. The
following matrix converts the eigenvectors coefficients into the
$(q,p)$ representation,
\begin{equation}
  \label{b7}
  \hat{A}=\left(\begin{array}{cccc}q_E&0&q_i&q_i\\0&g_0&p_i&-p_i\end{array}\right).
\end{equation}
The inverted matrix $\hat{A}^{-1}$ is given by
\begin{equation}
  \label{b8}
  \hat{A}^{-1}=\left(\begin{array}{cc}g_0&0\\0&q_E\\\frac{1}{2}p_i&\frac{1}{2}q_i\\\frac{1}{2}p_i&-\frac{1}{2}q_i\end{array}\right).
\end{equation}
The monodromy matrix in its eigenspace representation is given by,
\begin{equation}
  \label{b9}
  \hat{M}=\left(\begin{array}{cccc}1&0&0&0\\\frac{dT}{dE}&1&0&0\\0&0&\lambda_i&0\\0&0&0&\lambda_i^{-1}\end{array}\right).
\end{equation}
Then, the conversion from the initial momenta to the final coordinates
is given by the matrix $\hat{F}$,
\begin{equation}
  \label{b10}
  \hat{F}=\hat{K}(\tau,0)\hat{A}\hat{M}\hat{A}^{-1}\hat{K}^{-1}(\tau,0),
\end{equation}
whose expansion up to second order in $\tau$ is given by,
\begin{eqnarray}
  \label{b11}
  &&\hat{F}\simeq\hat{A}\hat{M}\hat{A}^{-1}
  \\
  \nonumber
  &&+\tau(\hat{K}_1\hat{A}\hat{M}\hat{A}^{-1}-\hat{A}\hat{M}\hat{A}^{-1}\hat{K}_1)
  \\
  \nonumber
  &&+ \frac{1}{2}\tau^2(\hat{K}_1^2A\hat{M}\hat{A}^{-1}+\hat{A}\hat{M}\hat{A}^{-1}\hat{K}_1^2-
     2 \hat{K}_1\hat{A}\hat{M}\hat{A}^{-1}\hat{K}_1)+...,
  \\
  \label{b12}
  &&\hat{K}_1=\left(\begin{array}{cc}0&I\\-\hat{k}&0\end{array}\right).
\end{eqnarray}

Now we turn to the calculation of the result of the application of
$\hat{F}$ to individual vectors. First we consider the trajectory with
initial momentum $g_0$. In zeroth order it is given by,
\begin{equation}
  \label{b14}
  \hat{A}\hat{M}\hat{A}^{-1}\left(\begin{array}{c}0\\g_0\end{array}\right)
  =\left(\begin{array}{c}0\\g_0\end{array}\right),
\end{equation}
which is the cause of the divergence of the determinant ratio at the
focal point, Eq.~\ref{ratio}. The first order terms are given
by,
\begin{equation}
  \label{b15}
  \hat{K}_1\hat{A}\hat{M}\hat{A}^{-1}\left(\begin{array}{c}0\\g_0\end{array}\right)
  =\hat{K}_1\left(\begin{array}{c}0\\g_0\end{array}\right)
  =\left(\begin{array}{c}g_0\\0\end{array}\right),
\end{equation}

\begin{eqnarray}
  \nonumber
  \hat{A}\hat{M}\hat{A}^{-1}\hat{K}_1\left(\begin{array}{c}0\\g_0\end{array}\right)
  &=&\hat{A}\hat{M}\hat{A}^{-1}\left(\begin{array}{c}g_0\\0\end{array}\right)
  \\
  \label{b17}
  &=&g_0^2\left(\begin{array}{c}q_E\\0\end{array}\right)+\frac{dT}{dE}g_0^2\left(\begin{array}{c}0\\g_0\end{array}\right)
  \\
  \nonumber
  &+&\frac{1}{2}\sum_i p_ig_0\left[(\lambda_i+\lambda_i^{-1})\left(\begin{array}{c}q_i\\0\end{array}\right)
  +(\lambda_i-\lambda_i^{-1})\left(\begin{array}{c}0\\p_i\end{array}\right)\right].
\end{eqnarray}
Collecting both terms, Eqs.~\ref{b15} and \ref{b17}, and taking into
account that
\begin{equation}
  \label{b18}
  g_0=g_0^2q_E+\sum_i(p_ig_0)q_i
\end{equation}
one obtains for the first order correction $q^{1st}_0$ to the
$q$-vector of the trajectory with initial momentum $g_0$,
\begin{equation}
  \label{b19}
  q^{1st}_0=\frac{\tau}{2}\sum_ip_ig_0(2-\lambda_i-\lambda_i^{-1})q_i,
\end{equation}
which does not contain any $q_i$-independent part. As a result, the
correction to $R$, Eq.~\ref{ratio}, disappears in the first order over
$\tau$.

We turn now to the calculation of the second order corrections to the
$q$-vector with respect to $\tau$ for the trajectory with initial
momentum $g_0$.  Only terms with a $q$-part that is linearly
independent of $q_i$ will contribute to the determinant in the
numerator of Eq.~\ref{ratio} to second order in $\tau$.  Both terms,
\begin{equation}
  \label{b20}
  \hat{K}_1^2\hat{A}\hat{M}\hat{A}^{-1}\left(\begin{array}{c}0\\g_0\end{array}\right)=
  \left(\begin{array}{c}0\\-\hat{k}g_0\end{array}\right)
\end{equation}
and
\begin{equation}
  \label{b25}
  \hat{A}\hat{M}\hat{A}^{-1}\hat{K}_1^2\left(\begin{array}{c}0\\g_0\end{array}\right)
 = \hat{A}\hat{M}\hat{A}^{-1} \left(\begin{array}{c}0\\-\hat{k}g_0\end{array}\right),                                                
\end{equation}
do not give a contribution to the $q$-part of the vector that is linearly independent of $q_i$.
The cross-term is given by,
\begin{eqnarray}
  \nonumber
  &&\hat{K}_1\hat{A}\hat{M}\hat{A}^{-1}\hat{K}_1\left(\begin{array}{c}0\\g_0\end{array}\right)=\hat{K}_1\left\{
  g_0^2\left(\begin{array}{c}q_E\\0\end{array}\right)+\frac{dT}{dE}g_0^2\left(\begin{array}{c}0\\g_0\end{array}\right)\right.
  \\
  \nonumber
  &&+\left.\frac{1}{2}\sum_i p_ig_0\left[(\lambda_i+\lambda_i^{-1})\left(\begin{array}{c}q_i\\0\end{array}\right)
  +(\lambda_i-\lambda_i^{-1})\left(\begin{array}{c}0\\p_i\end{array}\right)\right]\right\}
  \\
  \label{b22}
  &&=\frac{dT}{dE}g_0^2\left(\begin{array}{c}g_0\\0\end{array}\right)
  +\frac{1}{2}\sum_i p_ig_0 (\lambda_i-\lambda_i^{-1})\left(\begin{array}{c}p_i\\0\end{array}\right)
  + ...
\end{eqnarray}
where we did not include the momentum part of the final vector.  Using
the relation,
\begin{equation}
  \label{b23}
  p_i=(p_ig_0)q_E+\sum_j(p_ip_j)q_j,
\end{equation}
and taking into account Eq.~\ref{b18}, one obtains
the second order correction $q^{2nd}_0$ to the $q$-vector of the
trajectory with initial momentum $g_0$,
\begin{equation}
  \label{b28}
  q^{2nd}_0=-\tau^2\left[\frac{dT}{dE}g_0^2+\frac{1}{2g_0^2}\sum_i (p_ig_0)^2
  (\lambda_i-\lambda_i^{-1})\right]g_0+...
\end{equation}

Next we turn to the calculation of the $q$-vectors for the trajectories with
initial momenta $p_i$. Only the terms up to first order in $\tau$ will
contribute and only $q_i$-independent parts of the first order terms
should be considered. The zero-order term is given by
\begin{equation}
  \label{b29}
  \hat{A}\hat{M}\hat{A}^{-1}\left(\begin{array}{c}0\\p_i\end{array}\right)
  =\frac{1}{2}(\lambda_i+\lambda_i^{-1}) \left(\begin{array}{c}0\\p_i\end{array}\right)
  +\frac{1}{2}(\lambda_i-\lambda_i^{-1}) \left(\begin{array}{c}q_i\\0\end{array}\right).
\end{equation}
The first order terms are given by,
\begin{eqnarray}
  \nonumber
  \hat{K}_1\hat{A}\hat{M}\hat{A}^{-1}\left(\begin{array}{c}0\\p_i\end{array}\right)
  &=&\frac{1}{2}(\lambda_i+\lambda_i^{-1})\left(\begin{array}{c}p_i\\0\end{array}\right)+...
  \\
  \label{b31}
  &=&\frac{1}{2}(\lambda_i+\lambda_i^{-1})\frac{p_ig_0}{g_0^2}\left(\begin{array}{c}g_0\\0\end{array}\right)+...
  \\
  \label{b32}
  \hat{A}\hat{M}\hat{A}^{-1}\hat{K}_1\left(\begin{array}{c}0\\p_i\end{array}\right)
  &=&\hat{A}\hat{M}\hat{A}^{-1}\left(\begin{array}{c}p_i\\0\end{array}\right)
 =\frac{p_ig_0}{g_0^2}\left(\begin{array}{c}g_0\\0\end{array}\right)+...,
\end{eqnarray}
where we have again used  Eqs.~\ref{b18} and \ref{b23}. Collecting
the terms given by Eqs.~\ref{b31} and \ref{b32} one obtains that the first
order correction $q^{1st}_i$ to  the $q$-vector for the trajectory with
initial momentum $p_i$ is given by,
\begin{equation}
  \label{b33}
  q^{1st}_i=\frac{\tau}{2}(\lambda_i+\lambda_i^{-1}-2)\frac{p_ig_0}{g_0^2}g_0+...
\end{equation}

Substituting Eqs.~\ref{b19}, \ref{b28}, and \ref{b33} into
Eq.~\ref{ratio}, one obtains,
\begin{equation}
  \label{b34}
  R\simeq-\tau^2\left[\frac{dT}{dE}g_0^2+\frac{2}{g_0^2}\sum_i (p_ig_0)^2
    \frac{\lambda_i+\lambda_i^{-1} -
      2}{\lambda_i-\lambda_i^{-1}}\right]
  \frac{|g_0,q_i|}{|g_0,p_i|}\prod_i\frac{\lambda_i-\lambda_i^{-1}}{2}.
\end{equation}

\setcounter{equation}{0}
\section{Perturbation Theory for $U(T)$ in the Angular
  Momentum Conservation Case}
The derivation is completely analogous to the previous
section. Equations~\ref{b1}-\ref{b6} and \ref{b10} stay as they
are. Equations~\ref{b7}-\ref{b9} are modified as follows,
\begin{equation}
  \label{c7}
  \hat{A}=\left(\begin{array}{cccccc}q_E&0&q_\xi&0&q_i&q_i\\0&g_0&0&p_\xi&p_i&-p_i\end{array}\right).
\end{equation}
The inverted matrix $\hat{A}^{-1}$ is given by
\begin{equation}
  \label{c8}
  \hat{A}^{-1}=\left(\begin{array}{cc}g_0&0\\0&q_E\\p_\xi&0\\0&q_\xi
                       \\\frac{1}{2}p_i&\frac{1}{2}q_i\\\frac{1}{2}p_i&-\frac{1}{2}q_i\end{array}\right).
\end{equation}
The monodromy matrix in its eigenspace representation is given by,
\begin{equation}
  \label{c10}
  \hat{M}=\left(\begin{array}{cccccc}
                     1                 &0&0&0&0&0
                  \\\frac{dT}{dE}&1&0&0&0&0
                  \\ 0                &0&\hat{I}&\hat{C}&0&0
                  \\ 0                &0&0&\hat{I}&0&0
                  \\ 0&0&0&0&\hat{\lambda}&0
                  \\ 0&0&0&0&0&\hat{\lambda}^{-1}
                \end{array}\right),
\end{equation}

Relations~\ref{b18} and \ref{b23} are substituted by,
\begin{eqnarray}
  \label{c11}
  g_0&=&g_0^2q_E+\sum_i(p_ig_0)q_i+\sum_\xi(p_\xi g_0)q_\xi,
  \\
  \label{c12}
  p_i&=&(p_ig_0)q_E+\sum_j(p_ip_j)q_j+\sum_\xi(p_ip_\xi)q_\xi,
  \\
  \label{c13}
  p_\xi&=&(p_\xi g_0)q_E+\sum_j(p_\xi p_j)q_j+\sum_\eta(p_\xi p_\eta)q_\eta.
\end{eqnarray}
Let us consider the trajectory with initial momentum $g_0$.  For this
trajectory Eqs.~\ref{b14}-\ref{b15} hold.  The term
$\hat{A}\hat{M}\hat{A}^{-1}\hat{K}_1$ is given by,
\begin{eqnarray}
  \nonumber
  &&\hat{A}\hat{M}\hat{A}^{-1}\hat{K}_1\left(\begin{array}{c}0\\g_0\end{array}\right)
  =\hat{A}\hat{M}\hat{A}^{-1}\left(\begin{array}{c}g_0\\0\end{array}\right)
  \\
  \label{c15}
  &&=g_0^2\left(\begin{array}{c}q_E\\0\end{array}\right)+\frac{dT}{dE}g_0^2\left(\begin{array}{c}0\\g_0\end{array}\right)
  +\sum_\xi(p_\xi g_0)\left(\begin{array}{c}q_\xi\\0\end{array}\right)
  \\
  \nonumber
  &&+\frac{1}{2}\sum_i p_ig_0\left[(\lambda_i+\lambda_i^{-1})\left(\begin{array}{c}q_i\\0\end{array}\right)
  +(\lambda_i-\lambda_i^{-1})\left(\begin{array}{c}0\\p_i\end{array}\right)\right],
\end{eqnarray}
Substituting Eq.~\ref{c11} into Eq.~\ref{b15} one can see that
Eq.~\ref{b19} holds.

The $q$-part of the second order terms for the trajectory with
initial momentum $g_0$ should be linearly independent of $q_i$ and
$q_\xi$ because the linearly dependent terms have higher order
dependence on $\tau$. Therefore, the terms
$\hat{A}\hat{M}\hat{A}^{-1}\hat{K}^2_1$ and
$\hat{K}^2_1\hat{A}\hat{M}\hat{A}^{-1}$ again give no independent
contribution to the $q$-part of the trajectory with initial
momentum $g_0$. For the cross-term
$\hat{K}_1\hat{A}\hat{M}\hat{A}^{-1}\hat{K}_1$ Eq.~\ref{b22} stays and
so does the second order correction $q^{2nd}_0$ expression,
Eq.~\ref{b28}.

For the trajectories with initial momenta $p_i$ and $p_\xi$ only
the first order terms contribute to the second order term in
Eq.~\ref{ratio}. Also, only the $q$-parts that are linearly
independent of $q_i$ and $q_\xi$ will contribute to Eq.~\ref{ratio} to
the second order on $\tau$. For the trajectories with
initial momenta $p_i$ Eqs.~\ref{b29}-\ref{b32} remain the same and so
does the expression for the first order correction $q^{1st}_i$,
Eq.~\ref{b33}.

For the trajectories with initial momenta $p_\xi$, the zero order term
$\hat{A}\hat{M}\hat{A}^{-1}$ is given by
\begin{equation}
  \label{c16}
  \hat{A}\hat{M}\hat{A}^{-1}\left(\begin{array}{c}0\\p_\xi\end{array}\right)
  = \sum_{\eta}C_{\xi,\eta}\left(\begin{array}{c}q_\eta\\0\end{array}\right)+\left(\begin{array}{c}0\\p_\xi\end{array}\right).
 \end{equation}
The first order terms are given by,
\begin{eqnarray}
  \label{c17}
  \hat{K}_1\hat{A}\hat{M}\hat{A}^{-1}\left(\begin{array}{c}0\\p_\xi\end{array}\right)
  &=&\left(\begin{array}{c}p_\xi\\0\end{array}\right)+...
  =\frac{p_{\xi}g_0}{g_0^2}\left(\begin{array}{c}g_0\\0\end{array}\right)+...
  \\
  \label{c18}
  \hat{A}\hat{M}\hat{A}^{-1}\hat{K}_1\left(\begin{array}{c}0\\p_\xi\end{array}\right)
  &=&\hat{A}\hat{M}\hat{A}^{-1}\left(\begin{array}{c}p_\xi\\0\end{array}\right)
 =\frac{p_{\xi}g_0}{g_0^2}\left(\begin{array}{c}g_0\\0\end{array}\right)+...,
\end{eqnarray}
and one can see that they cancel each other. Thus, there is no first order
contribution to the $q$-vector from the trajectories with initial momenta $p_\xi$.

Collecting all the terms in Eq.~\ref{ratio} together, one obtains,
\begin{eqnarray}
  \label{c19}
  R&\simeq&-\tau^2\left[\frac{dT}{dE}g_0^2+\frac{2}{g_0^2}\sum_i (p_ig_0)^2
    \frac{\lambda_i+\lambda_i^{-1} -
      2}{\lambda_i-\lambda_i^{-1}}\right]
  \\
  \nonumber
  &\times&\frac{|g_0,q_i,\sum_{\eta}C_{\xi,\eta}q_{\eta}|}{|g_0,q_i,p_{\xi}|}\prod_i\frac{\lambda_i-\lambda_i^{-1}}{2}.
\end{eqnarray}

\setcounter{equation}{0}
\section{Entanglement Factor at Small Energies}

The potential is described by Eq.~\ref{multi_pot_exp} and only the
cubic anharmonicity is considered.  We start the trajectory from the
turning point with small energy $E$.  In the zeroth order
approximation the trajectory is given by,
\begin{eqnarray}
  \label{g2}
  q^{0th}_0(t)&=&-q_0\cos(\omega_bt)=-\frac{1}{2}q_0e^{i\omega_bt}+c.c.,\ \frac{1}{2}\omega_b^2q_0^2=E,
  \\
  \nonumber
  q^{0th}_i(t)&=&0.
\end{eqnarray}
The first order correction $q^{1st}_i(t)$ satisfies the following
equations,
\begin{eqnarray}
  \label{g3}
  \ddot{q}_0^{1st}+\omega_b^2q^{1st}_0&=&-\frac{1}{2}V_{0,0,0}^{(3)}q_0^{0th^2},
  \\
  \label{g4}
  \ddot{q}_i^{1st}-\omega_i^2q^{1st}_i&=&-\frac{1}{2}V_{0,0,i}^{(3)}q_0^{0th^2},  
\end{eqnarray}
The solution of Eqs.~\ref{g3}-\ref{g4} is given by,
\begin{eqnarray}
  \label{g5}
  q^{1st}_0(t)&=&-\frac{1}{4}\frac{q_0^2}{\omega_b^2}V_{0,0,0}^{(3)}(1-\frac{1}{3}\cos(2\omega_bt)),
  \\
  \label{g6}
  q^{1st}_i(t)&=&\frac{1}{4}q_0^2V_{0,0,i}^{(3)}(\omega_i^{-2}+\frac{1}{4\omega_b^2+\omega_i^2}\cos(2\omega_bt)).
\end{eqnarray}
Then the potential gradient $g_0=\partial V/\partial q$ at the
turning point is given by,
\begin{eqnarray}
  \label{g7}
  \frac{\partial V}{\partial q_0}&\simeq&-\omega_b^2q_0,
  \\
  \label{g8}
  \frac{\partial{V}}{\partial{q}_i}&\simeq&-\omega_i^2q_i^{1st}(0)+\frac{1}{2}V^{(3)}_{i,0,0}q_0^{0th}(0)^2
                                          =q_0^2V_{0,0,i}^{(3)}\frac{\omega_b^2}{4\omega_b^2+\omega_i^2}.                                 
\end{eqnarray}

We proceed with calculating the eigenvectors $v_{i,\pm}=(q_i,\pm p_i)$
of the monodromy transformaiton. To obtain the monodromy matrix one
can use Eqs.~\ref{ins_lem}.  To zeroth order the solutions of
these equations, which correspond to the eigenvectors
$v_{i,\pm}$, are given by,
\begin{eqnarray}
  \label{g9a}
  q^{0th}_{i,\pm,j}(t)=\delta_{i,j}q_i^0e^{\pm\omega_it},
  \\
  \label{g9b}
  q_i^0=\omega_i^{-1/2},
\end{eqnarray}
where Eq.~\ref{g9b} is the consequence of the orthogonality relation,
Eq.~\ref{ort_rel0}. To first order with respect to the potential
anharmonicity one can neglect the change in the instanton
trajectory. Then the first order correction to $q^{0th}_{i,+,j}(t)$
satisfies the following equations (the correction to
$q^{0th}_{i,-,j}(t)$ is obtained by changing the sign in the exponent,
so we omit the $\pm$ index in the notation),
\begin{eqnarray}
  \label{g10}
  \ddot{q}^{1st}_{i,0}+\omega_b^2q^{1st}_{i,0}=-V^{(3)}_{0,0,i}q_0^{0th}q_i^0e^{\omega_it},
  \\
  \label{g11}
  \ddot{q}^{1st}_{i,j}-\omega_j^2q^{1st}_{i,j}=-V^{(3)}_{0,i,j}q_0^{0th}q_i^0e^{\omega_it},
\end{eqnarray}
where $q_0^{0th}$ is given by Eq.~\ref{g2}. 
The solution of Eqs.~\ref{g10}-\ref{g11} is
given by,
\begin{eqnarray}
  \label{g12}
  q^{1st}_{i,0}(t)=\frac{1}{2}V^{(3)}_{0,0,i}q_0q_i^0\left(\frac{e^{(i\omega_b+\omega_i)t}}{(i\omega_b+\omega_i)^2+\omega_b^2}+c.c.\right),
  \\
  \label{g13}
  q^{1st}_{i,j}(t)=\frac{1}{2}V^{(3)}_{0,i,j}q_0q_i^0\left(\frac{e^{(i\omega_b+\omega_i)t}}{(i\omega_b+\omega_i)^2-\omega_j^2}+c.c.\right),
\end{eqnarray}
From Eqs.~\ref{g12}-\ref{g13} one can see that in the first order
approximation, the stability parameters $u_i=2\pi\omega_i/\omega_b$ do
not change and that $(q_i^{1st},\dot{q}_i^{1st})$ provide the first
order corrections to the corresponding eigenvectors.  We focus on
Eq.~\ref{g12}, which corresponds to the projection of the eigenvector
on the reaction coordinate, as only it contributes to leading
order. The first order correction to the momentum $p_i$ is given by
$\dot{q}_{i,0}^{1st}$,
\begin{eqnarray}
  \nonumber
  \dot{q}_{i,0}^{1st}&=&\frac{1}{2}V^{(3)}_{0,0,i}q_0q_i^0\left(\frac{i\omega_b+\omega_i}{\omega_i(i2\omega_b+\omega_i)}+c.c.\right)
  \\
  \label{g14}
                     &=&\frac{1}{2}V^{(3)}_{0,0,i}q_0q_i^0\left(\frac{1}{\omega_i}+\frac{\omega_i}{\omega_i^2+4\omega_b^2}\right).
\end{eqnarray}
The correction to the momentum changes sign when one considers the
conjugated eigenvector $v_{i,-}$, which corresponds to the decaying
exponent in Eq.~\ref{g9a} and negative sign for $\omega_i$ in
Eq.~\ref{g14}, as it should be. Using Eqs.~\ref{g7}, \ref{g8}, and
\ref{g14}, one obtains that $p_ig_0$ to leading order with respect to
energy is given by,
\begin{equation}
  \label{g15}
  p_ig_0=-2V^{(3)}_{0,0,i}q_0^2q_i^0\omega_b^4\omega_i^{-1}(\omega_i^2+4\omega_b^2)^{-1}.
\end{equation}
Substituting Eq.~\ref{g15} into Eq.~\ref{add_tun} and using
Eqs.~\ref{g7} and \ref{g9b}, one obtains Eq.~\ref{add_tun_cub}
approaches as the temperature approaches the crossover.

\setcounter{equation}{0}
\section{Action Anharmonicity $\alpha$ for Multi-Dimensional
  Potential}
To calculate $\alpha$ we use the relation
$\alpha=\frac{1}{2}\omega_b^2\frac{dT}{dE}|_{E=0}$. We start the
trajectory from the saddle point with small energy $E$ and then
calculate the anharmonic correction to the period of the harmonic
oscillator $T_0=2\pi/\omega_b$ with second order perturbation
theory. In the zeroth order approximation the trajectory is given by,
\begin{eqnarray}
  \label{e2}
  q^{0th}_0(t)&=&q_0\sin(\omega_bt),\ \frac{1}{2}\omega_b^2q_0^2=E,
  \\
  \nonumber
  q^{0th}_i(t)&=&0.
\end{eqnarray}
The first order correction $q^{1st}_i(t)$ satisfies the following
equations,
\begin{eqnarray}
  \label{e3}
  \ddot{q}_0^{1st}+\omega_b^2q^{1st}_0&=&-\frac{1}{2}V_{0,0,0}^{(3)}q_0^{0th^2},
  \\
  \label{e4}
  \ddot{q}_i^{1st}-\omega_i^2q^{1st}_i&=&-\frac{1}{2}V_{0,0,i}^{(3)}q_0^{0th^2},  
\end{eqnarray}
The solution of Eqs.~\ref{e3}-\ref{e4} is given by,
\begin{eqnarray}
  \label{e5}
  q^{1st}_0(t)&=&-\frac{1}{4}\frac{q_0^2}{\omega_b^2}V_{0,0,0}^{(3)}(1+\frac{1}{3}\cos(2\omega_bt)),
  \\
  \label{e6}
  q^{1st}_i(t)&=&\frac{1}{4}q_0^2V_{0,0,i}^{(3)}(\omega_i^{-2}-\frac{1}{4\omega_b^2+\omega_i^2}\cos(2\omega_bt)).
\end{eqnarray}
It is worth noting that, while the first order corrections,
Eqs.~\ref{e5}-\ref{e6}, modify the trajectory energy slightly, this
change is of the order of $~E^2$ or higher and, therefore, can be neglected.
From Eq.~\ref{e5} one can see that the period does not change in the
first order of perturbation theory. The equation for the second order correction
to the reaction coordinate $q_0^{2nd}$ reads,
\begin{equation}
  \label{eq7}
  \ddot{q}_0^{2nd}+\omega_b^2q^{2nd}_0=-V_{0,0,0}^{(3)}q_0^{0th}q_0^{1st}-\sum_iV_{0,0,i}^{(3)}q_0^{0th}q_i^{1st},
\end{equation}
whose solution is given by (we keep only the terms that change over the
period),
\begin{equation}
  \label{e8}
  q_0^{2nd}(t)=- t\cos(\omega_bt)\frac{1}{8}\left[\frac{5}{6}V_{0,0,0}^{(3)^2}\frac{q_0^3}{\omega_b^3}
    -\sum_iV_{0,0,i}^{(3)^2}\frac{q_0^3}{\omega_b}\left(\omega_i^{-2}
      +\frac{1}{2}\frac{1}{4\omega_b^2+\omega_i^2}\right)\right]+...
\end{equation}
Thus, the change in the reaction coordinate $\delta q_0$ during the full period
$2\pi/\omega_b$ is given by,
\begin{equation}
  \label{e9}
  \delta q_0=-\frac{\pi}{4}\left[\frac{5}{6}V_{0,0,0}^{(3)^2}\frac{q_0^3}{\omega_b^4}
    -\sum_iV_{0,0,i}^{(3)^2}\frac{q_0^3}{\omega_b^2}\left(\omega_i^{-2}
      +\frac{1}{2}\frac{1}{4\omega_b^2+\omega_i^2}\right)\right].
\end{equation}
The corresponding change in the period
$\delta T=-\delta q_0/\dot{q}_0^{0th}=-\delta q_0/(q_0\omega_b)$ and
$\alpha$ is given by Eq.~\ref{alpha_m3}.  For the potential with
fourth order anharmonicities, $\alpha$ is obtained in the first order
approximation in a similar fashion and is given by Eq.~\ref{alpha_m4}.

We proceed with the derivation of the multi-dimensional correction to
$\alpha$ in the MEP approximation, in which the tunneling path follows the
potential gradient. In the vicinity of the saddle point the deviation
from the $q_0$ direction happens mainly due to the $V^{(3)}_{0,0,i}$ term in the
potential and the MEP satisfies the following equation,
\begin{equation}
  \label{e10}
  \frac{dq_i}{dq_0}=\frac{1}{2}\frac{V^{(3)}_{0,0,i}}{\omega_b^2}q_0-\frac{\omega_i^2}{\omega_b^2}\frac{q_i}{q_0},
\end{equation}
whose solution is given by,
\begin{eqnarray}
  \label{e11a}
  q_i&=&\frac{1}{2}\frac{V^{(3)}_{0,0,i}}{2\omega_b^2+\omega_i^2}q_0^2,
  \\
  \label{e11b}
  \frac{dq_i}{dq_0}&=&\frac{V^{(3)}_{0,0,i}}{2\omega_b^2+\omega_i^2}q_0.
\end{eqnarray}
Then the abbreviated action, which is given by Eq.~\ref{short_act}, can be
written as,
\begin{eqnarray}
  \nonumber
  \tilde{S}(E)&=&\oint{d}q_0\sqrt{2E-\omega_b^2q_0^2+\sum_{i=1}(\omega_i^2q_i^2-V^{(3)}_{0,0,i}q_0^2q_i)-...}
                  \sqrt{1+\sum_{i=1}(dq_i/dq_0)^2}
  \\
  \nonumber
              &=&\frac{1}{8}\sum_{i=1}\frac{V^{(3)^2}_{0,0,i}}{(2\omega_b^2+\omega_i^2)^2}
                       \oint{d}q_0\left(4q_0^2\sqrt{2E-\omega_b^2q_0^2}-q_0^4\frac{4\omega_b^2+\omega_i^2}{\sqrt{2E-\omega_b^2q_0^2}}\right)+...
  \\
  \nonumber
              &=&-\frac{1}{8}\sum_{i=1}\frac{V^{(3)^2}_{0,0,i}}{\omega_b^2}\frac{8\omega_b^2+3\omega_i^2}{(2\omega_b^2+\omega_i^2)^2}
                  \oint{d}q_0q_0^2\sqrt{2E-\omega_b^2q_0^2}+...
  \\
  \label{e12}
              &=&-\frac{\pi}{8}E^2\sum_{i=1}\frac{V^{(3)^2}_{0,0,i}}{\omega_b^5}\frac{8\omega_b^2+3\omega_i^2}{(2\omega_b^2+\omega_i^2)^2}+...,
\end{eqnarray}
where we did not include one-dimensional and small terms. Then the
multi-dimensional correction to $\alpha$ in the MEP approximation is given
by Eq.~\ref{mep_alpha}.

\setcounter{equation}{0}
\section{The Relation between the One-Dimensional Potential and
  Abbreviated Action Expansions}
In this section we assume that the potential is written in the
following form,
\begin{equation}
  \label{one_pot_exp}
  V(q)=\frac{1}{2}\omega_b^2q^2+\frac{1}{6}V_3\omega_b^{5/2}q^3
  +\frac{1}{24}V_4\omega_b^3q^4+... .
\end{equation}
In mass-weighted coordinates $q$ has the dimensionality of
$[E]^{-1/2}$ and therefore the third and fourth order anharmonicity
parameters $V_3$ and $V_4$ are dimension-less.

The abbreviated action $\tilde{S}(E)$ corresponding to the potential
$V(q)$ can be written as,
\begin{equation}
  \label{act_com}
  \tilde{S}(E)=\oint dq\sqrt{2(E-V(q))},
\end{equation}
and the integral in Eq.~\ref{act_com} is considered in the complex
plane along a contour that is assumed to be large enough to
surround the turning points. Then Eq.~\ref{act_com} can be expanded
as,
\begin{eqnarray}
  \label{act_com1}
  \tilde{S}(E)&=&i\sqrt{2}\oint dq\sqrt{V(q)}\sqrt{1-E/V(q)}
  \\
  \nonumber
              &=&i\sqrt{2}\oint dq
  (\sqrt{V(q)}-\frac{1}{2}EV(q)^{-1/2}-\frac{1}{8}E^2V^{-3/2}+...).
\end{eqnarray}
Each sequential term corresponds to the next order in the expansion of $\tilde{S}(E)$
over $E$.  For the zero-order term one, of course, gets zero
contribution because the $\sqrt{V(q)}$ expansion contains only positive
powers of $q$.  Representing $\sqrt{V(q)}$ in the form,
\begin{eqnarray}
  \label{act_com2}
  \sqrt{V(q)}&=&2^{-1/2}\omega_bqF(q)^{1/2},
  \\
  \nonumber
  F(q)&=&1+\frac{1}{3}V_3\sqrt{\omega_b}q+\frac{1}{12}V_4\omega_bq^2+...,
\end{eqnarray}
one can see that to calculate the $n$-th order term one needs to find the
coefficient in front of $q^{2n-2}$ in the expansion of
$F(q)^{-(2n-1)/2}$.  Then the first order term is
$\frac{2\pi E}{\omega_b}$ ($\oint dq/q=2\pi i$) and the second order term
is $\pi(\frac{5}{24}V_3^2-\frac{1}{8}V_4)E^2/\omega_b^2$. Comparing with
Eq.~\ref{act_exp}, one arrives at Eq.~\ref{alpha_res}.

In the following we find a potential for which the action expansion in
Eq.~\ref{act_exp} is limited to the first two terms.  To this end we
transform Eq.~\ref{act_com} a bit differently, namely we introduce
another variable $y=\sqrt{V(q)}$ and use it as an independent
one. Because $V(q)$ is an analytic function of $q$, whose expansion
starts with a quadratic term, $q$ is also an analytic function of $y$
whose expansion starts with a linear term. The expression for
$\tilde{S}$ then reads,
\begin{equation}
  \label{act_mod}
\tilde{S}(E)=\sqrt{2}\oint dy\sqrt{E-y^2}\frac{dq}{dy},  
\end{equation}
Expanding $dq/dy=\sum_{n=0}^{\infty}A_ny^n$, one obtains,
\begin{equation}
  \label{act_mod1}
\tilde{S}(E)=\sqrt{2}\sum_{i=0}^{\infty}A_n\oint dy\sqrt{E-y^2}y^n,  
\end{equation}
Then we substitute $t=\sqrt{E}/y$ and take powers of $E$ out of the
integrals getting as a result,
\begin{equation}
  \label{act_mod2}
\tilde{S}(E)=i\sqrt{2}\sum_{i=0}^{\infty}A_nE^{n/2+1}\oint dt\sqrt{1-t^2}t^{-n-3},  
\end{equation}
Note that the integration contour now surrounds $t=0$ but not the
turning points. Expanding the function $\sqrt{1-t^2}$ one notes that
only even terms contribute to $\tilde{S}(E)$. We will assume that
$A_{2n+1}=0$. Then for the first two terms one obtains,
\begin{equation}
  \label{act_mod3}
  \tilde{S}(E)=\sqrt{2}\pi A_0E +2^{-3/2}\pi A_2E^2.
\end{equation}
Comparing with Eq.~\ref{act_exp} one finds that
$A_0=\sqrt{2}/\omega_b$ and $A_2=2^{3/2}\pi^{-1}\alpha/\omega_b^2$.
All other $A_n=0$. Now remembering that $A_n$ are expansion coefficients of
$dq/dy$ over powers of $y$, one obtains the following equation,
\begin{equation}
  \label{act_mod4}
  \frac{dq}{dy}=A_0+A_2y^2,
\end{equation}
whose solution is
\begin{equation}
  \label{act_mod5}
  q=A_0y+\frac{A_2}{3}y^3,
\end{equation}
where we have taken into account that $y=0$ at $q=0$. Substituting the
expressions for $A_0$ and $A_2$ and remembering that $y=\sqrt{V(q)}$,
one obtains the implicit equation for $V(q)$, Eq.~\ref{act_mod6}.

\setcounter{equation}{0}
\section{Tunneling Rate Constant at Low Temperatures for
  One-Dimensional Metastable Potential}
In this appendix we show that Langer's theory gives the correct
rate constant at very low temperatures, $\beta\omega_0\gg1$.  We
calculate the tunneling period $T(V^\#-E)$ and the abbreviated action
$\tilde{S}(V^\#-E)$ for the potential shown in Fig.~\ref{fig_pot} (we
use now the real potential, not the inverted one) at the energy $E$
close to the bottom of the well, which is now counted from the bottom
of the real well upward. We will assume that in the vicinity of the
potential minimum the potental energy is exactly quadratic,
\begin{equation}
  \label{f1}
  V(q) = \frac{1}{2}\omega_0^2q^2,\ q<q_2,
\end{equation}
where $\omega_0$ is the frequency at the bottom of the well. This
assumption, while it does not affect the final result, simplifies the
derivation. Then, the tunneling period at energy $E$ counted from the
bottom of the well can be written as,
\begin{equation}
  \label{f2}
  T(V^\#-E)=2\left(\int_{q_-}^{q_2}\frac{dq}{\sqrt{\omega_0^2q^2-2E}}+\int_{q_2}^{q_+}\frac{dq}{\sqrt{2(V(q)-E)}}\right),
\end{equation}
where $q_\pm$ are the turning points, see
Fig.~\ref{fig_pot}. Considering the expansion of the period $T$ over
energy $E$ at small
$E$ one finds that the second term gives some constant and higher
order terms $E^n,\ n>0$, while the first term gives
$2\ln(2q_2/q_-)/\omega_0$, where we have used the identity
\begin{equation}
  \label{f3}
  \int \frac{dq}{\sqrt{q^2-a^2}}=\ln\!\left[q/a+\sqrt{(q/a)^2-1}\right].
\end{equation}
Collecting the two terms and taking into account that $q_-=\sqrt{2E}/\omega_0$ one obtains,
\begin{equation}
  \label{f4}
  T(V^\#-E)= -\frac{1}{\omega_0}\ln(E/E_0) + ...,
\end{equation}
where the constant $E_0$ corresponds to the zero-order term
in the expansion of $T$ over $E$ and higher order terms are not
considered. Integrating Eq.~\ref{f4}, one arrives at Eq.~\ref{lt_act}. 

The tunneling rate constant at very low temperatures,
$\beta\omega_0\gg1$, corresponds to the decay of the ground state.
The tunneling rate constant for the metastable ground state wave
function $\psi(q)$ can be written as a probability flux $j$ to the
right of the barrier,\cite{landau3}
\begin{equation}
  \label{f5}
  k=\frac{i}{2}\left(\psi\frac{d\psi^*}{dq}-\psi^*\frac{d\psi}{dq}\right).
\end{equation}
Far to the right of the turning point $q_+$ the semiclassical approximation is
applicable and is given by,\cite{landau3}
\begin{eqnarray}
  \label{f6}
  &&\psi(q)\simeq C_+p^{-1/2}e^{i\int_{q_+}^qdq'p(q')},
  \\
  \nonumber
  &&p(q)=\sqrt{2(E-V(q))},\ q>q_+.
\end{eqnarray}
Substituting Eq.~\ref{f6} into Eq.~\ref{f5} one obtains that
$k=|C_+|^2$.

To calculate the normalization constant $C_+$ one needs to trace the
wave function $\psi(q)$ all the way to the well. According to the correspondence
rule the wave function to the left of the barrier is given
by\cite{landau3}
\begin{eqnarray}
  \label{f7}
  &&\psi(q)\simeq C_+p^{-1/2}e^{\int^{q_+}_qdq'p(q')},
  \\
  \nonumber
  &&p(q)=\sqrt{2(V(q)-E)},\ q<q_+,
\end{eqnarray}
where only the growing to the left exponent is considered. Tracing it
all the way to the left one can write an expression for $\psi(q)$ in
the vicinity of the left turning point $q_-$, to
the right of it,
\begin{eqnarray}
  \label{f8}
  &&\psi(q)\simeq C_+e^{\frac{1}{2}\tilde{S}(V^\#-E)}p^{-1/2}e^{-\int_{q_-}^qdq'p(q')},
  \\
  \nonumber
  &&p(q)=\sqrt{2(V(q)-E)},\ q_-<q,
\end{eqnarray}
Assuming that $E=\omega_0/2$ (the ground state) and using Eq.~\ref{f1} one can expand the
power of the exponent in $\exp(-\int_{q_-}^qdq'p(q'))$ as,
\begin{eqnarray}
  \label{f9}
  &&\int_{q_-}^qdq'p(q') =\omega_0\int_{q_-}^qdq'\sqrt{q'^2-\omega_0^{-1}}
  \\
  \nonumber
  &&\simeq\frac{1}{2}(\omega_0q^2-\ln(2e^{1/2}\omega_0^{1/2}q)),\ q_-=\omega_0^{-1/2},
\end{eqnarray}
where we have used the identity,
\begin{equation}
  \label{f10}
  \int dq\sqrt{q^2-a^2}=\frac{1}{2}\left(q\sqrt{q^2-a^2}-a^2\ln\!\left[q/a+\sqrt{(q/a)^2-1}\right]\right).
\end{equation}
Substituting Eq.~\ref{f9} into Eq.~\ref{f8} and comparing the result
with the ground wave function of the harmonic
oscillator,\cite{landau3}
$\psi(q)=(\omega_0/\pi)^{1/4}\exp(-\omega_0q^2/2)$, one obtains that,
\begin{equation}
  \label{f11}
  k=\frac{\omega_0}{2\pi^{1/2}e^{1/2}}e^{-\tilde{S}(V^\#-\omega_0/2)}
  =\frac{1}{\sqrt{2\pi}}\sqrt{\omega_0E_0}e^{-\tilde{S}^\#}
\end{equation}
where we have used Eq.~\ref{lt_act}.

On the other hand, in Langer's theory the rate constant is given by
Eqs.~\ref{tst} and \ref{one_deep}. From Eq.~\ref{f4} one finds that,
\begin{eqnarray}
  \label{f12}
  &&E=E_0e^{-\beta\omega_0},
  \\
  \label{f13}
  &&\Delta E= \sqrt{-\frac{dE}{d\beta}}=\sqrt{E_0\omega_0}e^{-\beta\omega_0/2},
\end{eqnarray}
where $\Delta E$ is the SPA energy integration window, which goes far beyond the
upper boundary of the energy integration when $\beta\omega_0\gg1$, cf. Eq.~\ref{f12}.
Substituting it into Eq.~\ref{one_deep} one obtains,
\begin{equation}
  \label{f14}
  Z^\#_{tun}=\sqrt{2\pi}\beta\sqrt{\omega_0E_0}e^{-\beta\omega_0/2}
  e^{-\tilde{S}^\#+\beta V^\#}.
\end{equation}
Substituting Eq.~\ref{f14} into Eq.~\ref{tst} and taking into account
that $Z_R\simeq e^{-\beta\omega_0/2}$ one arrives at Eq.~\ref{f11}.

\section*{Acknowledgments} 
This work was supported by the U. S. Department of Energy, Office of
Basic Energy Sciences, Division of Chemical Sciences, Geosciences, and
Biosciences under DOE Contract Number DE-AC02-06CH11357. YG was
supported as part of the Argonne-Sandia Consortium on High-Pressure
Combustion Chemistry, FWP 59044, while SJK was supported through the
Gas Phase Chemical Physics program.

\bibliography{../bib/vtst.bib,../bib/books.bib,comments.bib} 

\begin{mcitethebibliography}{63}
\bibitem[com()]{comment1}
In the separable case at low temperatures Eq.~2.29 in Ref.~\cite{miller75}
  reads $P(E)\simeq e^{-2\theta(E)-T(E)\sum_i\omega_i/2}$. For unimolecular
  decompositions $2\theta(E)$ and $T(E)=2d\theta/dE$ can be approximated by
  Eqs.~\ref{lt_act} and \ref{f4} of the present paper, correspondingly. It is
  easy to see that the energy dependence of the second term, which was
  neglected in Ref.~\cite{miller75} in the steepest descent approximation while
  deriving Eq.~2.34, will dominate at small $E$. For the Eckart potential
  $\theta(E)$ and $T(E)$ can be obtained from Eq.~\ref{eck1} of this paper and,
  again, the second term dominates near the bottom of the barrier.\relax
\mciteBstWouldAddEndPunctfalse
\mciteSetBstMidEndSepPunct{\mcitedefaultmidpunct}
{}{\mcitedefaultseppunct}\relax
\EndOfBibitem
\end{mcitethebibliography}
\end{document}